\documentclass[aps,pre,twocolumn,superscriptaddress,floatfix]{revtex4}

\usepackage{hyperref}
\usepackage{amsmath,amssymb}
\usepackage{float}
\usepackage{bm}
\usepackage{bbm}
\usepackage{graphicx, float, caption}

\usepackage{color}

\newcommand{\comebb}[1]{\textcolor{black}{#1}}
\newcommand{\addjg}[1]{\textcolor{black}{#1}}
\newcommand{\addeb}[1]{\textcolor{black}{#1}}

\newcommand{\addreview}[1]{\textcolor{black}{#1}}

\usepackage[normalem]{ulem}

\newcommand{\ddr}{\ensuremath{\mathrm{d}}}                                 
\newcommand{\deriv}[2]{\ensuremath{\frac{\mathrm{d} #1}{\mathrm{d} #2}}}   
\newcommand{\pderiv}[2]{\ensuremath{\frac{\partial #1 }{ \partial #2}}}    
\newcommand{\config}[0]{\ensuremath{\mathcal{C}}}

\newcommand{\order}[2]{\ensuremath{\mathcal{O}({#2}^{#1})}}

\newlength{\longme}

\begin{document}

\title{Nonequilibrium chemical potentials of steady-state lattice gas models in contact: \\
  A large-deviations approach}

\date{\today}

\author{Jules \surname{Guioth}}
\email{jules.guioth@damtp.cam.ac.uk}
\affiliation{DAMTP, Centre for Mathematical Sciences, University of Cambridge, Wilberforce Road, Cambridge CB3 0WA, UK}
\affiliation{Univ.~Grenoble Alpes, CNRS, LIPhy, F-38000 Grenoble, France}

\author{Eric \surname{Bertin}}
\email{eric.bertin@univ-grenoble-alpes.fr}
\affiliation{Univ.~Grenoble Alpes, CNRS, LIPhy, F-38000 Grenoble, France}

\begin{abstract}
We introduce a general framework to describe the stationary state of two driven systems exchanging particles or mass through a contact, in a slow exchange limit. The definition of chemical potentials for the systems in contact requires that the large-deviations function describing the repartition of mass between the two systems is additive, in the sense of being a sum of contributions from each system. We show that this additivity property \addreview{does not hold for an arbitrary contact dynamics, but} is satisfied on condition that a macroscopic detailed balance condition holds at contact, and that the coarse-grained contact dynamics satisfies a factorization property.
However, the nonequilibrium chemical potentials of the systems in contact keep track of the contact dynamics, and thus do not obey an equation of state.
These nonequilibrium chemical potentials can be related either to the equilibrium chemical potential, or to the nonequilibrium chemical potential of the isolated systems.
Results are applied both to an exactly solvable driven lattice gas model, and
to the Katz-Lebowitz-Spohn model using a numerical procedure to evaluate the chemical potential.
\addreview{The breaking of the additivity property is also illustrated on the exactly solvable model.}
\end{abstract}


\maketitle 

\section{Introduction}

The notion of intensive parameters conjugated to conserved quantities like energy, volume or number of particles lies at the very heart of equilibrium thermodynamics, yielding the key notions of temperature, pressure and chemical potentials that take equal values when two systems are put into contact.
A key issue in order to generalize thermodynamics to nonequilibrium steady states is to be able to define temperature, pressure and chemical potentials \cite{Oono1998,sasa2006steady}. In spite of many attempts, the notion of temperature in driven steady-state systems has eluded a thermodynamically consistent definition due to the lack of energy conservation \cite{Jou03,Cugliandolo11,Levine07,Bertin04,Martens09}.
However, conservation laws may still hold for volume and number of particles, so that it is natural to ask whether nonequilibrium pressure and chemical potential could be meaningfully defined in such systems.
A key feature such parameters should obey is that they should equalize when two systems in contact are able to exchange a globally conserved quantity like volume or particles. 
Nonequilibrium intensive thermodynamic parameters are also expected to fulfill a generalization of the zeroth law of thermodynamics. This means that if two systems have reached a steady state when separately put in contact with a third one, then they are also in steady state when brought into contact.
A related, but different, issue is that the thermodynamic parameters would be expected not to depend on the detailed way the two systems are put into contact, but only on bulk properties of each system. If this is the case, an equation of state holds. 
Although equations of state are generally present in equilibrium systems, their existence in steady-state driven systems is not granted, as shown by the generic lack of an equation of state for the mechanical pressure of gases of active particles \cite{solon2015nat}. For such systems, an equation of state is recovered, though, if specific symmetries are present \cite{solon2015prl}.

In the framework of lattice models of interacting driven particles, a nonequilibrium chemical potential has been defined under the hypothesis that an additivity condition is fulfilled \cite{bertin2006def,bertin2007intensive}.
This condition, not to be confused with the additivity condition  used to evaluate the current fluctuations in boundary driven systems \cite{bodineau2004}, states that if the system is decomposed into two subsystems $A$ and $B$, the large-deviations function $I(\rho_A)$ of the number $N_A$ of particles in subsystem $A$ can be written as a sum of two contributions, one depending only on $N_A$, and the other depending only on $N_B$:
\begin{equation} \label{eq:additivity}
P(N_A|N) \sim e^{-V [I_A(\rho_A) + I_B(\rho_B)]} \, ,
\end{equation}
where $V$ is the volume of the system, and $N=N_A+N_B$ is the fixed total number of particles; $\rho_{\alpha}=N_{\alpha}/V_{\alpha}$ is the density in subsystem $\alpha=A,B$.
\addreview{This property has already been considered some times ago in the context of the generic derivation of nonequilibrium hydrodynamic equations beyond local equilibrium for driven diffusive systems \cite{eyink1996hydrodynamics} 
Later, this additivity relation has been shown to be satisfied} for models like the zero range process \cite{evans2005nonequilibrium} and its continuous mass generalizations \cite{evans2004factorized}, where the $N$-body distribution factorizes in steady state.
Yet, even if a chemical potential can be defined in a single system by considering virtual partitions into subsystems, an important issue is whether this chemical potential predicts the steady-state density reached in two different systems in contact. Numerical simulations of lattice particle models like the Katz-Lebowitz-Spohn (KLS) model
\cite{pradhan2010nonequilibrium,pradhan2011approximate,dickman2014inconsistencies} as well as a lattice gas with nearest-neighbor exclusion
\cite{dickman2014inconsistencies,dickman2014failure} showed that depending on the dynamics of the contact, the steady-state density may or may not be correctly predicted. The validity of a nonequilibrium generalization of the zeroth law of thermodynamics has been verified numerically with a reasonable accuracy, but visible deviations have been reported \cite{pradhan2010nonequilibrium,pradhan2011approximate}.
The consequences of these results for the very existence of the notion of phase coexistence in nonequilibrium steady states have been emphasized \cite{dickman2016phase}.
The major role played by the dynamics of the contact has also been outlined using theoretical arguments or exact solutions of stochastic models
\cite{bertin2007intensive,pradhan2011approximate,chatterjee2015zeroth}.
To circumvent this difficulty, Sasa and Tasaki (ST) \cite{sasa2006steady} have proposed to use a specific type of contact dynamics modeling a high energy barrier between the two systems. This physical picture implies both a small exchange rate, thus implementing in practice the slow exchange concept in a nonequilibrium situation, and transfer rates from one system to the other that depend only on the configuration of the system from which particles are transferred. This class of transfer rates has been argued to play a key role in the phenomenological definition of chemical potentials proposed in \cite{sasa2006steady};
the consistency of this definition has been validated numerically in lattice particle models \cite{dickman2014inconsistencies}.
Note also that it has been proposed recently to define a subclass of contact dynamics for which the zeroth law remains valid by construction
\cite{chatterjee2015zeroth}.
However, the corresponding condition is not fulfilled in most realistic situations where the drive modifies the statistics of configurations in the system.

This paper aims at answering the following open main questions: (i) Can one link the phenomenological definition of chemical potential proposed by ST \cite{sasa2006steady} to the additivity condition of the large-deviations function of the number of particles in one system \cite{bertin2006def,bertin2007intensive,martens2011influence,pradhan2010nonequilibrium,pradhan2011approximate,chatterjee2015zeroth}?
(ii) Can one identify, going beyond the previously studied lattice particle models, the class of contact dynamics providing a consistent definition of chemical potential based on the additivity condition?
(iii) Does this chemical potential obey the zeroth law, and does it satisfy an equation of state? 
Focusing on the small exchange rate limit, we show that the additivity condition can be satisfied for a class of contact dynamics that is broader than the ST class, on condition that a macroscopic detailed balance relation holds. 
We discuss the issue of whether the chemical potential defined in systems in contact obeys an equation of state, and under which condition does the zeroth law holds. Note that a short and partial account of the present results has been published in \cite{guioth2018large}.

The paper is organized as follows.
Section~\ref{sec:large_dev} introduces the general framework of stochastic particle models in slow exchange limit at contact and the coarse-grained dynamics of densities.
Section~\ref{sec:large:dev} provides a large-deviations analysis of this coarse-grained density dynamics, and introduces the notion of macroscopic detailed balance.
Section~\ref{sec:additivity_large_dev} then discusses sufficient conditions for the large-deviations function of densities to be additive, thus allowing for the definition of nonequilibrium chemical potentials. The role of the contact dynamics in the properties of the chemical potentials is emphasized.
These definitions and properties are then illustrated in Sec.~\ref{sec:lattice_models} on the explicit example of an exactly solvable lattice gas model.
A numerical determination of the chemical potential in the (nonsolvable) KLS model is also presented. 
Finally, Sec.~\ref{sec:discussion} discusses how the present results may shed light on previously reported (and sometimes puzzling) results.

\section{General framework}
\label{sec:large_dev}




\subsection{Stochastic driven lattice models}
\label{sec:def_dyn}

We begin with the general definition of the models considered. Stochastic lattice gases and mass transport models are formally continuous time stochastic Markovian systems defined on lattice \cite{liggett2012interacting,spitzer1970} composed of interacting particles that jump from site to site. One can describe microscopic state or configuration by the occupation number in each site $x$ of the lattice. 

Examples of such lattice models are the well known asymmetric simple exclusion process (ASEP) as well as its variants \cite{spitzer1970,derrida2007non,derrida1998asep}, the Katz-Lebowitz-Spohn (KLS) model \cite{katz1984nonequilibrium,zia2010twenty}, the zero range process \cite{evans2005nonequilibrium,Levine2005}, as well as its numerous variants \cite{Evans2006,evans2004factorized,evans2006factorized,zia2004construction}, etc.

For one system, we note $\Lambda \subset \mathbb{Z}^{\mathrm{d}}$ the space grid ($d$ being the space dimension), $V=|\Lambda|$ the number of sites, $N$ the number of particles and $\config = \{ n_{x} \}_{x \in \Lambda} $ a configuration of the system, $n_{x}\in [0, n_{\mathrm{max}}]$ being the number of particles at site $x$ ($n_{\mathrm{max}}$ can be finite or infinite). The local configuration $n_{x}$ is generically an integer for most models, but it can be a real variable $n_{x} \geq 0$ in some models where it has been called a ``continuous mass'' \cite{Evans2006,evans2004factorized,evans2006factorized,zia2004construction}. We point out that periodic boundary conditions are assumed at least in the drive direction (details are given below).

The dynamics is entirely prescribed by the transition rates $T(\config^{\prime}|\config)$ to jump from a configuration $\config$ to another one $\config^{\prime}$. For instance, for stochastic lattice gases, $\config'$ corresponds to a single move of one particle from a site to another.
As these simple models intend to be mesoscopic modelings of the dynamics of particles, one imposes the \emph{local detailed balance} \cite{katz1984nonequilibrium,maes2003origin,maes2003time} condition which restricts the class of systems that can be modeled by Markov processes. It states that
\begin{equation}
  \label{eq:local_DB}
  \frac{T(\config^{\prime}|\config)}{T(\config|\config^{\prime})} = \exp \left[ -\beta \left(E(\config^{\prime})-E(\config) - W(\config,\config^{\prime}) \right) \right]
\end{equation}
where $E(\config)$ is the energy of the configuration $\config$ and $W(\config,\config^{\prime})$ refers to the nonconservative work associated with the drive.
Physically, the local detailed balance assumption means that the underlying heat bath stays in equilibrium at inverse temperature $\beta$ despite the force applied on the particles (see introduction of \cite{wynants2010structures}).

  The energy $E$ is generically prescribed by a given interaction potential sometimes supplemented by an external potential. The nonconservative work $W$ depends on the drive but we will mostly consider a constant driving force $\bm{f}$ ($\|\bm{f}\| = f$) for which
  \begin{equation*}
    W(\config,\config^{\prime})= \bm{f}\cdot{}\bm{j}(\config,\config^{\prime}),
  \end{equation*}
  $\bm{j}(\config,\config^{\prime})$ being the total current flowing in the system for the transition $\config \to \config^{\prime}$ (the latter is thus generally localized if only one particle jumps at a time). We note that the explicit functional form of transition rates $T(\config^{\prime}|\config)$ is not completely specified by the local detailed balance property. We will consider in specific examples below some common choices obeying local detailed balance such as the exponential rule, the Kawasaki rule, the Metropolis rule or the Sasa-Tasaki rule \cite{tasaki2004remark}.

\subsection{Contact dynamics between two systems}
\label{sec:def_contact}

We define in this subsection the contact dynamics between two systems $A$ and $B$ defined by their own Hamiltonians $E_{A}(\config_{A})$, $E_{B}(\config_{B})$ and their own driving force $f_{A}$, $f_{B}$. One calls $\Lambda_{k}$ the space grid of system $k$, $V_{k}=|\Lambda_{k}|$ the number of sites of system $k$ and $N_{k}=\mathcal{N}(\config_{k})$ the actual number of particles in system $k$, $k=A,B$. One sets $\gamma_{A}=V_{A}/V$ and $\gamma_{B}=V_{B}/V$ ($\gamma_{A}+\gamma_{B}=1$) the relative sizes of system $A$ and $B$ with respect to the total volume
$V=V_{A}+V_{B}$. The contact dynamics is defined through a transition rate $T_{\mathrm{c}}(\config_{A}^{\prime},\config_{B}^{\prime}|\config_{A},\config_{B})$ obeying local detailed balance as well. The total number of particles $N=\mathcal{N}(\config_{A})+\mathcal{N}(\config_{B})$ is assumed to be fixed. \addreview{We assume in the following that both $A$ and $B$ are in contact with heat baths at the same inverse temperature $\beta$. One should note however, that taking into account different heat baths at different (inverse) temperatures $\beta_{A}$ and $\beta_{B}$ is feasible in general (see Sec.~\ref{sec:micro_trans_rate_factorization} for some more details).}

As mentioned in the Introduction, our main goal is to investigate the situation of two uniform nonequilibrium systems in contact. Since we have chosen to look at the simple situation of externally driven systems for which periodic boundary conditions along the driving forces are necessary, the natural contact geometry one can think of is an \emph{orthogonal} contact to the driving forces $f_{A}$ and $f_{B}$. Hence, microscopic transition rates at contact, $T_{c}$, are assumed not to depend on driving forces $f_{A}$, $f_{B}$. The case with an additional dependence on the forcing at contact will be briefly discussed later
in Sec.~\ref{sec:extra-work_tilt}.

The dynamics of the whole system composed of systems $A$ and $B$ is thus prescribed by transition rates in the bulk as well as the contact ones. The stochastic process is a Poisson Markov jump process and the probability to observe a configuration $\config~=~(\config_{A},\config_{B})$ at time $t$, $P_{t}(\config)$, obeys the following master equation
\begin{align}
  \label{eq:micro_master_equation_composed_system_AB}
  & \deriv{P_{t}}{t}(\config_{A},\config_{B})   \\
  & = \hspace{1em} \sum_{\config_{A}'\neq \config_{A}} T_{A}(\config_{A}|\config_{A}') P_{t}(\config_{A}',\config_{B}) - \lambda_{A}(\config_{A})P_{t}(\config_{A},\config_{B})  \notag \\
  & \hspace{1.5em} + \sum_{\config_{B}' \neq \config_{B}} T_{B}(\config_{B}|\config_{B}') P_{t}(\config_{A}, \config_{B}') - \lambda_{B}(\config_{B})P_{t}(\config_{A},\config_{B}) \notag \\
  & \hspace{1.5em} + \sum_{\substack{\config_{A}' \neq \config_{A} \\ \config_{B}' \neq \config_{B} }} T_{c}(\config_{A},\config_{B}|\config_{A}',\config_{B}') P_{t}(\config_{A}',\config_{B}') \notag \\
  & \hspace{12em} - \lambda_{c}(\config_{A},\config_{B}) P_{t}(\config_{A},\config_{B}) \; \notag
\end{align}
with $\lambda_{k}(\config)=\sum_{\config'\neq\config}T_{k}(\config'|\config)$ the escape rates associated with the configuration $\config$, $k=A$, $B$, or $c$.

\subsection{Coarse-grained dynamics of the densities}

Our goal is to compute the stationary distribution of the number of particles in each system, knowing the total number of particles $N=N_{A}+N_{B}$ or rather the density $\bar{\rho}=\gamma_{A}\rho_{A}+\gamma_{B}\rho_{B}$. If microscopic detailed balance holds, one can solve straightforwardly the stationary master equation \eqref{eq:micro_master_equation_composed_system_AB} and thus derive directly the distribution of densities $\rho_{A},\, \rho_{B}$ in each system. However, since both systems are out-of-equilibrium, detailed balance does not hold. The strategy is then to derive an evolution equation on the probability distribution on $\rho_{A}$.

One can easily derive an evolution equation on the probability $P_{t}(\rho_{A}|\bar{\rho})$ to observe a density $\rho_{A}= N_{A}/V_{A}$ (and $\rho_{B}=\gamma_{B}^{-1}(\bar{\rho}-\gamma_{A}\rho_{A})$ since mass is conserved), by summing over all the microstates $\config=(\config_{A},\config_{B})$ corresponding to the given density in \eqref{eq:micro_master_equation_composed_system_AB}. Since the dynamics in the bulks of $A$ and $B$ conserve the number of particles in each system, the coarse-grained master equation over $(\rho_{A},\rho_{B})$ only involves the dynamics at contact encoded in $T_{c}$. It yields
\begin{align}
  \label{eq:coarse-grained_master_eq_formal}
  &\deriv{P_{t}}{t}(\rho_{A}|\bar{\rho})  \\
  & \quad = \sum_{\rho_{A}' \neq \rho_{A}} \pi_{\bar{\rho}, \, t}(\rho_{A}|\rho_{A}') P_{t}(\rho_{A}'|\bar{\rho}) - \pi_{\bar{\rho} ,\, t}(\rho_{A}'|\rho_{A}) P_{t}(\rho_{A}|\bar{\rho}) \notag .
\end{align}
The quantity $\pi_{\bar{\rho},\, t}(\rho_{A}'|\rho_{A})$ refers to the coarse-grained transition rate associated with the coarse-grained transition $\rho_{A}\to \rho_{A}'=\rho_{A}-\Delta N_{A}/V_{A}$. It reads as
\begin{align}
  \label{eq:coarse_grained_transrate_time}
  &\pi_{\bar{\rho} , \,  t}(\rho_{A}'|\rho_{A})  \\
  &  \quad =\sideset{}{_{c}^{(\Delta N_{A})}}\sum_{\config_{A}', \config_{B}'} \sideset{}{_{c}^{(0)}} \sum_{\config_{A}, \config_{B}} T_{c}(\config_{A}', \config_{B}'|\config_{A}, \config_{B}) P_{t}(\config_{A}, \config_{B}|\rho_{A}, \bar{\rho}) \notag.
\end{align}
with $\sum_{c}^{(\Delta N_{A})}$ referring to the sum over configurations $\config_{A}', \config_{B}'$ that, respectively, contain $\mathcal{N}(\config_{A}')=\rho_{A}V_{A}+\Delta N_{A}$ and $\mathcal{N}(\config_{B}')=\rho_{B}V_{B}-\Delta N_{A}$ particles (the second sum being exactly the same with $\Delta N_{A} = 0$).

The knowledge of the coarse-grained transition rates thus rests upon the knowledge of the conditional probability distributions $P_{t}(\config_{A},\config_{B}|\rho_{A},\bar{\rho})$ whose coupled evolutions can be obtained from the microscopic dynamics \eqref{eq:micro_master_equation_composed_system_AB}. However, the latter is not tractable in general.
We identify in the next subsections a limit in which the probability distribution $P_{t}(\config_{A}, \config_{B}|\rho_{A}, \bar{\rho})$ can be evaluated, in order to determine the coarse-grained transition rate $\pi_{\bar{\rho}, \, t}(\rho_{A}'|\rho_{A})$.

But, before that, one first needs to deal with another limit, namely, the thermodynamic limit, and thus to specify the volume dependence of the coarse-grained transition rates $\pi_{\bar{\rho}, \, t}(\rho_{A}'|\rho_{A})$ defined in \eqref{eq:coarse_grained_transrate_time}.

\subsection{Volume dependence of the macroscopic transition rates at contact}

The transition rate $\pi_{\bar{\rho}, t}(\rho_{A}'|\rho_{A})$ is associated with the following transition:
\begin{equation}
  \label{eq:detailed_transition}
  \begin{split}
  \rho_{A}= \frac{N_{A}}{V_{A}} & \rightarrow \rho_{A}'= \frac{N_{A} + \Delta N_{A}}{V_{A}} \\
  \rho_{B}= \frac{N_{B}}{V_{B}} & \rightarrow \rho_{B}'= \frac{N_{B} - \Delta N_{A}}{V_{B}} .
  \end{split}
\end{equation}
In all this work we naturally assume that the number of particles that can be exchanged per unit time (during a single transition) is bounded and does not scale with the volume of the system. We then define
 \begin{equation}
  \label{eq:starting_point_coarse-grained_transition_rate}
  \pi_{\bar{\rho}, t}(\rho_{A}'|\rho_{A}) \equiv \nu(V) \varphi_{V, t}(\rho_{A}, \rho_{B} ; \Delta N_{A})
\end{equation}
 where we have explicitly introduced the $\rho_{B}$ dependence as well as the volume $V$ which refers to the possible volume dependence of the transition rate (according to notations introduced before, $V_{A}= \gamma_{A} V$ and $V_{B}=\gamma_{B}V$, with $\gamma_{A}, \, \gamma_{B}$ kept finite at the thermodynamic limit). The $V$-dependence of the transition rate is potentially twofold.
 The first contribution, encoded in the factor $\nu(V)$, models the contact area and how it grows as $V\to\infty$.
 The second contribution, which has to be considered case by case may appear through the probability distribution $P_{t}(\config_{A}, \config_{B}|\rho_{A}, \rho_{B})$ as a potential finite-size effect. If the number of sites that connect both systems is fixed, then $\varphi_{V, t}$ is not proportional to $V$ and $\nu(V)=\nu$. The remaining $V$ dependence in $\varphi_{V, t}$ is expected to vanish as $V\to\infty$ so that $\lim_{V\to\infty} \varphi_{V, t}$ exists. In this case, the dynamics for a large system size is slower than the one at small system size, and this $V$ dependence may be absorbed in the time scale, as discussed below.
However, if the contact area grows with volumes, the frequency factor $\nu(V)$ is expected to be proportional to $V^{\alpha}$, $\alpha\leqslant 1$ (for instance, if the contact is proportional to the external area, then $\alpha=1-1/d$, $d$ being the space dimension), in addition to potential finite-size contributions.
This $V$ dependence will be discussed explicitly in specific systems,
but we assume in what follows that the main dependence on the volume $V$ is included in the factor $\nu(V)$, and that
$\varphi=\lim_{V\to\infty} \varphi_{V, t}$ is well defined.


\section{Large-deviations analysis of the density dynamics}
\label{sec:large:dev}

The study of the thermodynamic limit $V \to \infty$ for a jump stochastic processes is reminiscent of the expansion of the master equation popularized by Van Kampen \cite{van1992stochastic}. Nevertheless, as stressed in the Introduction, a thermodynamic analysis based on stochastic dynamics requires a large-deviations analysis that is not captured by the Van Kampen expansion (at least when truncating the expansion at a finite order).
Even if we are not interested in rare events \emph{per se}, the large-deviations framework is the relevant one to study the dominant extensive contribution to the probability distribution of density $\rho_{A}$ (and $\rho_{B}$), exactly as it is for equilibrium statistical mechanics (see for instance \cite{touchette2009large}). One should note that this large-deviations analysis, on the same kind of master equations as considered by Van Kampen, was first considered, with a somewhat different emphasis, in \cite{kubo1973fluctuation} (see also \cite{maes2007static}). Also, even if the work presented here has been developed independently, we should mention the recent study of Ge and Qian \cite{ge2017mathematical} which deals with the same kind of large-deviations analysis in the context of chemical reactions. 

The simplest way (even though not rigorous) to look at a large-deviations scaling is to introduce the large-deviations ansatz directly in the nonhomogeneous master equation \eqref{eq:coarse-grained_master_eq_formal}. To treat systems $A$ and $B$ on the same footing, we introduce
\begin{equation}
  \label{eq:large_dev_ansatz}
  P_{t}(\rho_{A}|\bar{\rho}) = P_{t}(\rho_{A}, \rho_{B} | \bar{\rho}) \asymp e^{-V \mathcal{I}_{t}(\rho_{A}, \rho_{B}|\bar{\rho})} \; ,
\end{equation}
where $\asymp$ refers to a logarithmic equivalence \footnote{Precisely, $\mathcal{I}_{t}~=~\lim_{V\to\infty} \ln (P_{t})/V $.} for large $V=V_{A}+V_{B}$. It yields
\begin{align}
  \label{eq:large_dev_analysis_cg_master_eq}
  & V\deriv{\mathcal{I}_{t}}{t}(\rho_{A},\rho_{B}|\bar{\rho}) \\
  & \hspace{0.8em} = \nu(V) \sum_{\Delta N_{A}} \varphi_{V, t}(\rho_{A}, \rho_{B} ; \Delta N_{A}) \notag\\
  & \hspace{1.9em} \times \left[ \exp\left\{ \Delta N_{A} \left( \gamma_{A}^{-1}\pderiv{\mathcal{I}_{t}}{\rho_{A}} - \gamma_{B}^{-1}\pderiv{\mathcal{I}_{t}}{\rho_{B}} \right) \right\} - 1 \right] \notag \\
  & \hspace{1.9em} + \order{-1}{V} . \notag
\end{align}
To go further, one needs to specify the time dependence of the coarse-grained transition rates $\varphi_{V, t}(\rho_{A}, \rho_{B} ; \Delta N_{A})$ in order to take the thermodynamic limit $V\to\infty$.

\subsection{Slow exchange limit at contact}

As stressed before, this time dependence of the coarse-grained transition rates refers in fact to the relaxation of the conditional probability distributions $P_{t}(\config_{A},\config_{B}|\rho_{A},\bar{\rho})$.
Let us introduce $\tau_{c} \sim V \nu(V)^{-1} \epsilon^{-1}$, where $\left[\nu(V) \epsilon \right]^{-1}$ is the typical time scale between two jumps of particles across the contact, $\epsilon$ being the typical value of the rates $T_{c}(\config_{A}', \config_{B}'|\config_{A}, \config_{B})$. To make this scaling even more explicit, we rewrite $T_{c}\to \epsilon T_{c}$ where $T_{c}$ is now of order $1$. The second time scale, called $\tau_{b}(V)$, is the one at which the bulks of each systems relax, which generally depends on the volumes.
If both time scales are not separated, the contact does induce a perturbation, at least locally, in each system, at odds with equilibrium systems where the detailed balance condition ensures the absence of perturbation. In driven systems, long-range correlations along the flux are rather ubiquitous \cite{bertini2007long,garrido1990long,spohn1983long,dorfman1994generic}, and a local perturbation may even produce long-range effects \cite{maes2009nonequilibrium}, leading to a strong coupling between systems.
Even if these long-range effects are less expected to happen when the contact is local and the extension in the directions perpendicular to the driving forces is large enough, this coupling remains too difficult to be studied in a general setting. Following the phenomenological study of \cite{sasa2006steady}, as well as their detailed study on the KLS model \cite{hayashi2003thermo,sasa2006steady}, we will then focus on the simpler situation for which the dynamics at contact is much slower than the dynamics in the bulk, meaning that $\tau_{c}$ is much larger than $\tau_{b}$. We will see that this limiting case is also the more likely to enable a thermodynamic structure since the stationary probability density happens to be almost factorized.

From a physical viewpoint, this low frequency exchange limit can be reached either by a high energy barrier that screens the interactions between both systems at contact, or by a low opening rate of a gate or strong conformation selection of particles that decreases the attempt rate of jumps without screening the interactions between systems in contact. Whatever the situation, one will consider in the following that the contact and the bulk time scales are well separated, meaning $\tau_{c} \gg \tau_{b}$. One will thus enforce explicitly $\tau_{b}/\tau_{c} \sim \tau_{b}(V) V^{-1} \nu(V) \epsilon \ll 1$ by tuning $\epsilon$ accordingly. In particular, in the $V\to\infty$ limit, one will need $\epsilon$ to decrease faster than $V \left[ \tau_{b}(V)\nu(V)\right ]^{-1}$.
Note that the relaxation time of the perturbation in the bulk, $\tau_{b}$ might actually not depend on $V$ if the effect of the perturbation is well localized, thus reducing the threshold at which the small exchange rates limit tends to be valid. In particular, if the width of the contact area is fixed, $\nu(V) = \nu$, and the time scale separation is satisfied when $\epsilon$ is small enough, but independent of $V$.
Hence, for $\epsilon$ small enough, jumps of particles between $A$ and $B$ are typically separated by large time intervals of typical length $\tau_{c}$ during which the bulks are mostly in their stationary states. At a resolution time very large compared to the bulk time $\tau_{b}$ (but smaller than $\tau_{c}$), the coarse-grained transition rate [Eq.~\eqref{eq:coarse_grained_transrate_time}] reads, at zeroth order in $\epsilon$, as
\begin{align}
  \label{eq:coarse_grained_transrate_rescaled_time}
  & \varphi_{V}(\rho_{A}'|\rho_{A})  \\
  &  \quad = \sideset{}{_{c}^{(\Delta N_{A})}}\sum_{\config_{A}', \config_{B}'} \sideset{}{_{c}^{(0)}} \sum_{\config_{A}, \config_{B}}  T_{c}(\config_{A}',\config_{B}'|\config_{A},\config_{B}) \notag \\
  & \qquad \qquad \qquad \qquad \qquad \times P_{A}(\config_{A}|\rho_{A}) P_{B}(\config_{B}|\rho_{B})  \; . \notag
\end{align}

One can recognize that the averaging is performed with respect to the stationary solution of the master equation \eqref{eq:micro_master_equation_composed_system_AB} without contact, namely, $T_{c}=0$. It is equal to the stationary distribution one would reach if the systems were completely isolated from each other, which is completely factorized: 
\begin{equation} \label{eq:factorized:dist}
P(\config_{A},\config_{B}|\rho_{A},\bar{\rho}) = P_{A}(\config_{A}|\rho_{A}) P_{B}(\config_{B}|\rho_{B}).
\end{equation}
\addreview{Several comments are in order here. First,}
note that microscopic detailed balance can still be broken at contact even in the low exchange rate limit, because the steady-state distributions of the two systems in contact are generically different from the equilibrium ones when the drives are switched on.
\addreview{Second, it is important to note that the factorization property (\ref{eq:factorized:dist}) of the joint distribution of microscopic configurations is valid for distribution conditioned to a given density of particles in each system ---this property results from the assumed time scale separation.
The factorization property may not be valid for the full (i.e., unconditioned) distribution of microscopic configurations, that can be written in the form
\begin{equation}
P(\config_{A},\config_{B}) = \int \ddr\rho_{A} P(\rho_A|\bar{\rho}) \, P_{A}(\config_{A}|\rho_{A}) P_{B}(\config_{B}|\rho_{B}),
\end{equation}
where we have assumed the validity of Eq.~(\ref{eq:factorized:dist}).
The density distribution $P(\rho_A|\bar{\rho})$ may not be factorized with respect to the two systems, even if the conditioned distribution $P(\config_{A},\config_{B}|\rho_{A},\bar{\rho})$ is factorized according to Eq.~(\ref{eq:factorized:dist}). We show below that the distribution $P(\rho_A|\bar{\rho})$ is determined by the coarse-grained dynamics at contact, the latter being determined under the factorization assumption (\ref{eq:factorized:dist}).
This is a key difference with Ref.~\cite{chatterjee2015zeroth}, where the unconditioned distribution of microscopic configurations, $P(\config_{A},\config_{B})$ is assumed to take a factorized form, which straightforwardly implies a factorized form of the density distribution $P(\rho_A|\bar{\rho})$.}
As a last remark, one should note that for finite $\epsilon$, such that the typical exchange time is of the order of the time for both systems to relax to their respective steady state, one can intuitively guess that an approximate description of the dynamics of densities involves relaxation modes of the bulk dynamics \cite{wang2016entropy,wang2016noneq}.
This much more complicated situation goes beyond the scope of this work, and will not be considered in this paper.

\subsection{Evolution equation of the large-deviations function of densities}

Eventually, in the slow exchange limit detailed above, and after having rescaled time $t \to V \left[ \epsilon \nu(V)\right]^{-1} t$, one obtains an equation on the large-deviations function $\mathcal{I}_{t}$ that reads as, in the thermodynamic limit $V\to\infty$,
\begin{multline}
   \label{eq:hamilton-jacobi_eq}
  \deriv{\mathcal{I}_{t}}{t}(\rho_{A},\rho_{B}|\bar{\rho}) = 
  \sum_{\Delta N_{A}} \varphi(\rho_{A},\rho_{B} ; \Delta N_{A}) \\
  \times \left[ \exp\left\{ \Delta N_{A} \left( \gamma_{A}^{-1}\pderiv{\mathcal{I}_{t}}{\rho_{A}} - \gamma_{B}^{-1}\pderiv{\mathcal{I}_{t}}{\rho_{B}} \right) \right\} - 1 \right] ,
\end{multline}
The latter equation generally bears the name of a Hamilton-Jacobi equation \cite{maes2007static}. In the absence of phase transition, the large-deviations function $\mathcal{I}$ is expected to be convex and to display only a single minimum characterized by the vanishing of the derivative of $\mathcal{I}$ \cite{touchette2009large,ellis2007entropy}.

The introduction at this point of the explicit $\rho_{B}$-dependence allows one to see more clearly the dependence on the relative sizes of the systems as well as the parallel with the situation at equilibrium that we remind one of here very briefly. Indeed, at equilibrium, the large-deviations function $\mathcal{I}$ is closely linked to the free energies of both systems, up to a temperature factor $\beta$. One has
\begin{align}
  \label{eq:equilibrium_large_dev}
  &\mathcal{I}_{\mathrm{eq}}(\rho_{A},\rho_{B}|\bar{\rho})  \\
  &\quad = \beta\gamma_{A}\left[f_{A}(\rho_{A}) - f_{A}(\rho_{A}^{\ast})\right] + \beta\gamma_{B}\left[f_{B}(\rho_{B})-f_{B}(\rho_{B}^{\ast})\right] , \notag
\end{align}
where $f_{k}$ refers to the equilibrium free energies per unit volume of system $k$ and $\rho_{k}^{\ast}$ the most probable density of system $k$ (which corresponds to the average density). The most probable densities are fixed by the vanishing of the derivative of $\mathcal{I}$ which reads as $f_{A}'(\rho_{A}^{\ast}) = f_{B}'(\rho_{B}^{\ast})$ (or, in other words, that chemical potentials defined as the derivative of the free energies, are equal).

Nevertheless, in order to lighten notations, we will come back from now on to our former convention and omit the $\rho_{B}$ dependence, which will be implicitly assumed through mass conservation, and simply write
\begin{equation}
  I_{t}(\rho_{A}|\bar{\rho}) = \gamma_{A}^{-1}\mathcal{I}_{t}(\rho_{A},\rho_{B}|\bar{\rho}) \,.
\end{equation}
This implies
\begin{equation}
  I_{t}'(\rho_{A}|\bar{\rho})=\gamma_{A}^{-1}\pderiv{\mathcal{I}_{t}}{\rho_{A}} - \gamma_{B}^{-1}\pderiv{\mathcal{I}_{t}}{\rho_{B}} \,,
\end{equation}
where the prime symbol indicates a derivative with respect to $\rho_{A}$. In this way,
\begin{equation}
  \label{eq:large_deviation_rho_A_only}
  P_{t}(\rho_{A}|\bar{\rho}) \asymp e^{-V_{A}I_{t}(\rho_{A}|\bar{\rho})} \; .
\end{equation}
With this notation, the Hamilton-Jacobi equation \eqref{eq:hamilton-jacobi_eq} simply reads as
\begin{equation}
  \label{eq:hamilton_jacobi_eq_rhoA}
  \deriv{I_{t}}{t}(\rho_{A}|\bar{\rho}) = \sum_{\Delta N_{A}\neq 0} \varphi(\rho_{A} ; \Delta N_{A}) \left[e^{\Delta N_{A}I_{t}'(\rho_{A}|\bar{\rho})} - 1 \right] .
\end{equation}
Assuming ergodicity, the stationary solution $I=\lim_{t\to\infty}I_{t}$ thus obeys, for all $\rho_A$,
\begin{equation}
  \label{eq:stationary_HJ_eq_rhoA}
  \sum_{\Delta N_{A}\neq 0} \varphi(\rho_{A} ; \Delta N_{A}) \left[e^{\Delta N_{A}I'(\rho_{A}|\bar{\rho})} - 1 \right] = 0 \; .
\end{equation}
In the following, we mostly use the notation $I(\rho_{A}|\bar{\rho})$, but we come back to the more explicit notation $\mathcal{I}(\rho_{A},\rho_{B}|\bar{\rho})$ when needed.

\subsection{Macroscopic detailed balance}
\label{sec:macro_detailed_balance}

To start at a formal level, one can notice that the Hamilton-Jacobi equation \eqref{eq:stationary_HJ_eq_rhoA} can be easily solved if each term under the rearranged sum cancels one by one for any $\rho_{A}$:
\begin{align}
  & \sum_{\Delta N_{A}\neq 0} \varphi(\rho_{A},\Delta N_{A})\left[ e^{\Delta N_{A} I'(\rho_{A}|\bar{\rho})} - 1 \right] \\
  & \hspace{1.5em} = \sum_{\Delta N_{A}\neq 0}\underbrace{\left[\varphi(\rho_{A},\Delta N_{A}) e^{\Delta N_{A} I'(\rho_{A}|\bar{\rho})} - \varphi(\rho_{A},-\Delta N_{A})\right]}_{=0\; \text{if detailed balance}} \notag \\
  & \hspace{1.5em}= 0 . \notag
\end{align}
One gets a generalized detailed balance condition, that we will call \emph{macroscopic detailed balance} in the following. It reads as
\begin{equation}
  \label{eq:macro_detailed_balance}
  I^{\prime}(\rho_{A}|\bar{\rho}) = \frac{1}{\Delta N_{A}} \ln \frac{\varphi(\rho_{A},-\Delta N_{A})}{\varphi(\rho_{A},\Delta N_{A})} \; .
\end{equation}
Importantly, one can note that for most lattice gas models, that deal with the dynamics of particles on lattice in continuous time (and potentially more realistic systems), only one particle can be exchanged at the same time. Thus, $\Delta N_{A} = \pm 1$ at most and one can easily check that the macroscopic detailed balance condition is always verified.
However, for more general situations when several particles can be simultaneously exchanged (or when $\Delta N_{A}$ corresponds to the exchange of a continuous mass), the macroscopic detailed balance condition is generically not fulfilled.
\addreview{This relation \eqref{eq:macro_detailed_balance} has already been considered in the literature discussing the existence of nonequilibrium chemical potentials and especially in \cite{chatterjee2015zeroth}.
However, let us emphasize here that the spirit of our present work is different from that of \cite{chatterjee2015zeroth}. In the latter, conditions that should be satisfied by the contact dynamics in order for the large-deviations function $\mathcal{I}(\rho_A,\rho_B)$ to be additive are identified. This defines how the contact dynamics should be fine tuned when varying the drives so that additivity remains satisfied.
In contrast, we fix the contact dynamics by assuming that it does not depend on the drive and satisfies microscopic detailed balance at equilibrium, and we then check whether the additivity property of $\mathcal{I}(\rho_A,\rho_B)$ still holds when switching on the drives.
In addition, we also emphasize that this macroscopic detailed balance relation is not linked to any microscopic detailed balance relation since the latter is generally broken as soon as the stationary distributions of each nonequilibrium isolated systems differ from the equilibrium ones and as the dynamics at contact is orthogonal to the driving forces. Also, the large-deviations function $I(\rho_{A}|\bar{\rho})$ is not directly attached to the distribution of the isolated systems as it is in \cite{chatterjee2015zeroth}. This natural connection only exists through the transition rates $\varphi(\rho_{A}, \Delta N_{A})$ that involve the stationary probability distributions of the isolated systems. Furthermore, Eq.~\eqref{eq:macro_detailed_balance} is not restricted to short-range correlated systems. The only assumption, shared with the study reported in \cite{chatterjee2015zeroth}, is the slow exchange limit.}
Eventually, one should note that this condition is an asymptotic consequence (in the thermodynamic limit) of an underlying time-reversal symmetry of the coarse-grained dynamics \cite{ge2017mathematical}.

When the macroscopic detailed balance is broken, the Hamilton-Jacobi equation \eqref{eq:stationary_HJ_eq_rhoA} has to be solved as a whole and finding a general solution is in general not reachable. Nevertheless, one can always try to find its solution perturbatively with respect to a known reference solution, often the equilibrium one. This will be detailed in a future publication \cite{GuiothTBP19}.

\subsection{Link between the $I'(\rho_{A} | \bar{\rho})$ and the current $J(\rho_{A})$}

Assuming that there is only one stationary state in the thermodynamic limit, the latter is naturally defined by the vanishing of the particle current $J(\rho_{A}^{\ast})=0$ through the contact. This deterministic current is defined in the infinite volume limit through the deterministic relaxation equation of the density $\rho_{A}$ which reads as
  \begin{equation}
    \label{eq:def:determinist_current}
    \deriv{\rho_{A}(t)}{t} = J(\rho_{A}(t)) = \sum_{\Delta N_{A}} \varphi(\rho_{A}, \Delta N_{A}) \, \Delta N_{A}  \; .
  \end{equation}
Of course, the characterization of the stationary state by the vanishing of the current should be consistent with the minimization of the large-deviations function $I(\rho_{A}|\bar{\rho})$ at $\rho_{A}= \rho_{A}^{\ast}$. In other words, the property $J(\rho_{A}^{\ast}) = 0$ has to be equivalent to $I'(\rho_{A}|\bar{\rho})=0$. This intuitive property can be shown by using the Hamilton-Jacobi equation \eqref{eq:stationary_HJ_eq_rhoA} \cite{ge2017mathematical}. We reproduce the argument in Appendix \ref{sec:appA}.

More interestingly, one can also show a link between the current $J(\rho_{A})$ and the derivative of $I$, $I'(\rho_{A}|\bar{\rho})$, understood as a thermodynamic force. To do so, one should first note that any transition rate $\varphi(\rho_{A} ; \Delta N_{A})$ can be decomposed in terms of a work (or thermodynamic force) $F(\rho_{A}, \Delta N_{A})$ to perform the transition $\rho_{A} \to \rho_{A} + \Delta N_{A}/V_{A}$, antisymmetric with respect to $\Delta N_{A}$, and a mobility factor $a(\rho_{A}, \Delta N_{A})$, symmetric with respect to $\Delta N_{A}$ \cite{maes2007and}:
  \begin{equation}
    \label{eq:decomposition_force_act_transrate}
    \varphi(\rho_{A}, \Delta N_{A}) = a(\rho_{A}, \Delta N_{A}) \, e^{ \tfrac{1}{2} F(\rho_{A}, \Delta N_{A}) } \, ,
  \end{equation}
  where $a(\rho_{A},\Delta N_{A}) = a(\rho_{A}, - \Delta N_{A})$ and $F(\rho_{A}, \Delta N_{A}) = -F(\rho_{A}, - \Delta N_{A})$.

  If macroscopic detailed balance holds, the work $F(\rho_{A}, \Delta N_{A}) = - I'(\rho_{A}|\bar{\rho}) \Delta N_{A}$ and thus
  \begin{equation}
    \label{eq:current_Iprime_MDB}
    \deriv{\rho_{A}(t)}{t} = - 2 \sum_{\Delta N_{A}>0} \Delta N_{A} a \sinh\left( \frac{I' \Delta N_{A}}{2} \right) \, . 
  \end{equation}
  If macroscopic detailed balance does not hold, one can introduce $F_{A}(\rho_{A}, \Delta N_{A})$ such that $F = - I'\Delta N_{A} + F_{A}$. The current then reads as
  \begin{align}
    \label{eq:current_Iprime_broken_MDB}
    & \deriv{\rho_{A}(t)}{t} =  \\
    & \hspace{2em} - 2 \sum_{\Delta N_{A}>0} \Delta N_{A} a\cosh\left( \tfrac{F_{A}}{2} \right) \sinh\left( \tfrac{I'\Delta N_{A}}{2} \right) \notag \\
    & \hspace{2em} + 2 \sum_{\Delta N_{A}>0} \Delta N_{A} a \cosh\left( \tfrac{I'\Delta N_{A}}{2}  \right) \sinh\left( \tfrac{F_{A}}{2} \right) \; . \notag 
  \end{align}
  The argument $(\rho_{A}, \Delta N_{A})$ in $F_{A}$, $I'$ and $a$ has been implicitly assumed in the two last equations.

 Relations between currents and thermodynamic forces in both equations \eqref{eq:current_Iprime_MDB} and \eqref{eq:current_Iprime_broken_MDB} are the nonlinear analogs of the linear flux (or force) relationship (linear response theory) in near-equilibrium irreversible thermodynamics \cite{degrootmazur1984, bertini2015macroscopic}.
\addreview{
   Note, however, that the entropy production at the thermodynamic limit is still expressed as a product of a particle flux and a thermodynamic force, $\dot{\mathcal{S}}=-JI'$ when $F_{A}=0$ \emph{i.e.} when macroscopic detailed balance holds. More generally, when $F_{A}\neq 0$, the total entropy production reads as $\dot{\mathcal{S}} = \sum_{\Delta N_{A}} \varphi(\rho_{A}, \Delta N_{A}) F(\rho_{A}, \Delta N_{A})$ \cite{ge2017mathematical}.
}
  
\section{Additivity property of the large-deviations function}
\label{sec:additivity_large_dev}

We now address the issue of the additivity of the large-deviations function $I(\rho_{A}|\bar{\rho})$ for two systems in contact. This additivity condition is reminiscent of the additivity of the free energy for equilibrium systems interacting through short ranged potentials, recalled in Eq.~\eqref{eq:equilibrium_large_dev}. It reads as 
\begin{equation}
  \label{eq:additivity_LDF}
  \mathcal{I}(\rho_{A},\rho_{B}|\bar{\rho}) \equiv \gamma_{A}I(\rho_{A}|\bar{\rho}) = \gamma_{A}I_{A}(\rho_{A}) + \gamma_{B}I_{B}(\rho_{B})
\end{equation}
where $\rho_{B}= \gamma_{B}^{-1}(\rho-\gamma_{A}\rho_{A})$. If such additivity condition holds, the derivative of the large-deviations function reads as
\begin{equation}
  \label{eq:chemical_pot_general_def_additivity}
  I'(\rho_{A}|\bar{\rho}) = I_{A}'(\rho_{A})  - I_{B}'(\rho_{B}) \; ,
\end{equation}
and the steady-state densities $\rho_{A}^{\ast}$ and $\rho_{B}^{\ast}$ satisfy $I_{A}'(\rho_{A}^{\ast})=I_{B}'(\rho_{B}^{\ast})$. Hence, it offers the possibility to attach to each system a quantity, $I_{k}'(\rho_{k})$ ($k=A\, ,B$), rather denoted as $\mu_{k}(\rho_{k})$ henceforth, that will be called \emph{generalised chemical potential} at contact.



In the following subsections, we first identify sufficient conditions in order for the large-deviations function $I(\rho_{A}|\bar{\rho})$ to be additive. Then, assuming that the additivity condition holds, we discuss expressions and properties of the chemical potentials thus defined. In particular, we make connection with chemical potentials of isolated systems and discuss the zeroth law of thermodynamics.

\addreview{When the identified conditions are not met, it is likely that the additivity property of the large-deviations function $I(\rho_{A}|\bar{\rho})$ no longer holds. We will briefly discuss this absence of additivity on particular cases in Sec.~\ref{sec:lattice_models}.}

\subsection{Chemical potential of systems in contact}
\label{sec:chempot}

\subsubsection{Factorization condition of the contact dynamics}

When macroscopic detailed balance \eqref{eq:macro_detailed_balance} holds, the additivity property of the large-deviations function $I(\rho_{A}|\bar{\rho})$ should be directly related to the coarse-grained transition rates $\varphi$.
\addreview{This implies that the ratio $\varphi(\rho_A,-\Delta N_A)/\varphi(\rho_A,\Delta N_A)$ should take a factorized form with respect to systems $A$ and $B$. A sufficient condition for this factorization condition to hold is to assume  that the coarse-grained transition rate factorizes as}
\begin{equation}
  \label{eq:factorization_transrate}
  \varphi(\rho_{A},\Delta N_{A}) = \nu_{0}\phi_{A}(\rho_{A},\Delta N_{A})\phi_{B}(\rho_{B},\Delta N_{B})
\end{equation}
with $\Delta N_{B} = -\Delta N_{A}$ and $\nu_{0}$ an arbitrary common frequency scale. The macroscopic detailed balance \eqref{eq:macro_detailed_balance} then enables one to split the derivative of the large-deviations function into two contributions that, respectively, depend on each systems $k=A,\, B$. It reads as
\begin{equation}
  I'(\rho_{A}|\bar{\rho}) = \mu_{A}^{\mathrm{cont}}(\rho_{A}) - \mu_{B}^{\mathrm{cont}}(\rho_{B}) \, ,
\end{equation}
where the chemical potentials are given by
\begin{equation}
  \label{eq:def:chempot_contact}
  \mu_{k}^{\mathrm{cont}}(\rho_{k}) \equiv \ln \frac{\phi_{k}(\rho_{k},-1)}{\phi_{k}(\rho_{k},1)}
\end{equation}
with $k=A,B$. \footnote{The expression (\ref{eq:def:chempot_contact}) of the chemical potential may be reminiscent of the interpretation, \emph{at equilibrium}, of the fugacity $\zeta = e^{\mu}$ as the ``probability to escape'' of a randomly chosen particle \cite[p. 77]{sekimoto2010stochastic}. However, in out-of-equilibrium systems, the lack of microscopic detailed balance is expected to break this escape probability interpretation.}
One notes that we have set $|\Delta N_{A}|=1$ since the large-deviations function $I(\rho_{A}|\bar{\rho})$ given by the macroscopic detailed balance condition \eqref{eq:macro_detailed_balance} does not depend on $\Delta N_{A}$.  At the most probable values of the densities $\rho_{A}^{\ast}$, $\rho_{B}^{\ast}$ (around which the probability density $P(\rho_{A}|\bar{\rho})$ is more and more peaked when system sizes increase), $I'(\rho_{A}^{\ast}|\bar{\rho}) = 0$, resulting in the equalization of the chemical potentials:
\begin{equation}
  \label{eq:equalization_chempot}
  \mu_{A}^{\mathrm{cont}}(\rho_{A}^{\ast}) = \mu_{B}^{\mathrm{cont}}(\rho_{B}^{\ast}) \, .
\end{equation}

\addreview{The factorization condition (\ref{eq:factorization_transrate}) of the contact dynamics is only a sufficient condition, and is \emph{a priori} not necessary. One could in principle imagine nonfactorized forms of the coarse-grained rate
$\varphi(\rho_{A},\Delta N_{A})$ such that the ratio $\varphi(\rho_A,-\Delta N_A)/\varphi(\rho_A,\Delta N_A)$ is factorized.
However, we will see below that the factorized form (\ref{eq:factorization_transrate}) is particularly relevant when trying to link the coarse-grained transition rates at contact to the corresponding microscopic transition rates.}

\subsubsection{Microscopic transition rates: Factorization condition}
\label{sec:micro_trans_rate_factorization}

We now relate the factorization assumption \eqref{eq:factorization_transrate} of the coarse-grained transition rates to the properties of the microscopic transition rates $T_{c}$.
As seen in Eq.~(\ref{eq:coarse_grained_transrate_rescaled_time}), the transition rates $\varphi(\rho_{A}, \Delta N_{A})$ are averages of the microscopic transition rates over the product of stationary distributions of the isolated systems. One can then observe that a sufficient condition is simply that the microscopic transition rates factorize in a similar way as the macroscopic ones
\begin{equation}
  \label{eq:factorization_micro_transrates}
  T_{c}(\config_{A}^{\prime},\config_{B}'| \config_{A},\config_{B}) = \nu_{0} \theta_{A}(\config_{A},\config_{A}^{\prime}) \theta_{B}(\config_{B},\config_{B}^{\prime}) \; .
\end{equation}
\addreview{Quite importantly, the factorized form (\ref{eq:factorization_transrate}) of the coarse-grained rates $\varphi(\rho_{A},\Delta N_{A})$ is obtained from the corresponding factorized form (\ref{eq:factorization_micro_transrates}) of the microscopic transition rates $T_{c}(\config_{A}^{\prime},\config_{B}'| \config_{A},\config_{B})$ for any form of the steady-state distributions $P_A(\config_{A})$ and $P_B(\config_{B})$.
In contrast, the situation is different for nonfactorized rates.
Let us assume that we know microscopic transition rates $T_{c}(\config_{A}^{\prime},\config_{B}'| \config_{A},\config_{B})$ that do not factorize, but are such that the ratio $\varphi(\rho_A,-\Delta N_A)/\varphi(\rho_A,\Delta N_A)$ is factorized for given steady-state distributions $P_A(\config_{A})$ and $P_B(\config_{B})$.
Then, changing these distributions $P_A(\config_{A})$ and $P_B(\config_{B})$ (by considering different systems, or simply by changing the value of the drive) generically breaks the factorization of the ratio $\varphi(\rho_A,-\Delta N_A)/\varphi(\rho_A,\Delta N_A)$, except if microscopic detailed balance hold,
as in \cite{chatterjee2015zeroth}.
The factorized rates (\ref{eq:factorization_micro_transrates}) thus appear much easier to handle, due to their robustness with respect to coarse graining. In the following, we shall restrict ourselves to the case of factorized rates when considering additive large-deviations functions.}

The local detailed balance assumption, discussed in the Introduction, imposes a constraint on the ratio between transition rates of a transition $\config\to\config'$ and its time-reversed counterpart $\config' \to \config$: the logarithm of the latter ratio is related to ($\beta$ times) the energy supplied by the environment (from operator and heat bath) to make the transition.
Yet, the local detailed balance hypothesis does not entirely define transition rates and we assume more generally that the latter only depends on ($\beta$ times) the work necessary to make the transition. One notes that this assumption is consistent with most of the common choices of transition rates present in the literature \cite{katz1984nonequilibrium,tasaki2004remark}. It reads as
\begin{align}
  \label{eq:LDB_contact_micro_transrate}
  T_{c}(\config'|\config) & = \tau\left( \beta W(\config,\config') \right) \\
                          & = \tau\left(\beta W_{A}(\config_{A},\config_{A}') + \beta W_{B}(\config_{B},\config_{B}') \right. \notag \\
                          & \hspace{5em} \left. + \beta W_{AB}^{\mathrm{int}}(\config_{A},\config_{A}' \, ; \, \config_{B}, \config_{B}') \right) , \notag
\end{align}
with  $W(\config,\config')$ the supplied work which has been split in several contributions which depend on systems $A$ and $B$ as well as their interactions. According to the local detailed balance condition, the function $\tau(x)$ should satisfy $\tau(x)=e^{x}\tau(-x)$. Clearly, the presence of the interaction term which mixes $A$ and $B$ configurations does not allow the above factorization property to hold in general. One should thus assume that the latter is negligible with respect to the other contributions. Among the classical choices that satisfy local detailed balance (\emph{e.g.} the exponential rule, the Metropolis rule, the Kawasaki/heat-bath rule and the Sasa-Tasaki rule, \emph{etc.}), only two of them verify the factorization condition \eqref{eq:factorization_micro_transrates}, as discussed below.

\addreview{As a remark, we briefly discuss the case for which systems $A$ and $B$ are in contact with independent thermostats at different (inverse) temperatures $\beta_{A}$ and $\beta_{B}$, respectively. First, we note that such a change does not break the slow exchange limit assumption: the stationary probability distributions of the isolated systems $P_{A}(\config_{A}|\rho_{A})$ and $P_{B}(\config_{B}|\rho_{B})$ would just depend on $\beta_{A}$ and $\beta_{B}$, respectively. As for the contact dynamics, this temperature inhomogeneity may be more challenging to take into account since one needs to know with which thermostat the energy is exchanged during the transition in order to estimate the entropy flux from the system toward the heat baths (local detailed balance assumption). If there is no interaction term $W_{AB}^{\mathrm{int}}$, it appears natural to assume that $T_{c}(\config' | \config) = \tau \left( \beta_{A} W_{A}(\config_{A},\config_{A}') + \beta_{B}W_{B}(\config_{B},\config_{B}') \right)$. In this case, $\beta_{A}$ and $\beta_{B}$ can be absorbed into a redefinition of the coupling parameters of $A$ and $B$, respectively.}

\subsubsection{Sasa-Tasaki dynamics}

The first one, discussed in \cite[Appendix B]{sasa2006steady}, will be called the Sasa-Tasaki rule.\footnote{Nevertheless, this choice of dynamics has been considered for long time. According to \cite[p. 112]{sekimoto2010stochastic}, the latter has already been considered in \cite{bergmann1955new,lebowitz1957irreversible}.} This rule is claimed to model a high energy barrier separating systems $A$ and $B$. If the energy barrier is high, the transition rate takes an Arrhenius expression:
\begin{equation}
  \label{eq:def:ST_transrate}
    T_{c}(\mathcal{C}_{A}', \mathcal{C}_{B}'|\mathcal{C}_{A},\mathcal{C}_{B}) = \epsilon
  \begin{cases}
    e^{-\beta\Delta H_{A}} \text{ if } \Delta N_{A}=-1 \\
    e^{-\beta\Delta H_{B}} \text{ if } \Delta N_{A}=+1 
  \end{cases}
  \, ,
\end{equation}
$\Delta H_{k}=H_{k}(\mathcal{C}_{k}')-H_{k}(\mathcal{C}_{k})$ and $\Delta N_{A}=\mathcal{N}(\mathcal{C}_{A}')-\mathcal{N}(\mathcal{C}_{A})$, $k=A,\, B$, $H_{A,B}$ being the respective energies of systems $A$ and $B$. We point out that mass conservation $\mathcal{N}(\mathcal{C}_{A}')-\mathcal{N}(\mathcal{C}_{A})=-(\mathcal{N}(\mathcal{C}_{B}')-\mathcal{N}(\mathcal{C}_{B}))$ is implicitly enforced in Eq.~\eqref{eq:def:ST_transrate}. Also, $\epsilon=e^{-\beta \Delta V} $ where $\Delta V$ is the height of the energy barrier separating $A$ and $B$. When the barrier is high, $\epsilon \ll 1$ and one gets a natural realization of the low frequency exchange limit.

\subsubsection{Exponential rule}

Another classic rule for which the factorization condition holds is when $\tau(x)=e^{x/2}$. It reads as
\begin{equation}
  \label{eq:def:exponential_rule_contact}
  T_{c}(\config_{A}',\config_{B}'|\config_{A},\config_{B}) = \epsilon \, e^{-\tfrac{\beta}{2} \Delta H_{A}} e^{-\tfrac{\beta}{2} \Delta H_{B}} \, ,
\end{equation}
where one has used the same notations as for the Sasa-Tasaki dynamics.
This exponential rule could be relevant in the case when interactions between $A$ and $B$ are negligible compared to the interactions within each system but when the slow exchange contact is generated by a conformation selection or low frequency openings of a gate but not by any high energy barrier.

Even if these two cases appear to be the most common transition rates that satisfy the factorization property \eqref{eq:factorization_micro_transrates}, one could certainly imagine other rules that might be relevant. That being said, we will nevertheless discuss the implication of the factorization property \eqref{eq:factorization_micro_transrates} in its general formulation without any reference to a specific choice, unless mentioned otherwise.

\subsection{Validity of the zeroth law of thermodynamics}
\label{sec:zeroth_law}

The zeroth law lies at the heart of equilibrium thermodynamics and deals with the issue of the contact between equilibrium systems.
It is sometimes referred to as a transitivity property of equilibrium states, meaning that if two systems $A$ and $B$ are in equilibrium with a third one $C$, they are in equilibrium with each other. If it holds, it can serve as an operational definition of the existence of intensive thermodynamic parameters related to exchange of conserved quantities through the contact, like temperature (for the exchange of energy), pressure (for the exchange of volume), or chemical potentials (for the exchange of particles), that equalize when systems are in equilibrium with each other. 

For contact between nonequilibrium driven systems in steady-state, one has seen the importance of the contact dynamics as it exerts a strong influence on the stationary densities in each system. In certain situations when macroscopic detailed balance holds and when macroscopic transition rates factorize, one can define intensive thermodynamic parameters, namely, chemical potentials, that are associated with each system and equalize when the stationary state is reached. However, this does not necessarily lead to the zeroth law as stated above since the chemical potential defined here may depend on the specificity of the contact between systems. For instance, if $A$ and $B$ are separately in contact with $C$ through different contact dynamics, it is not at all guaranteed that the final stationary states of $A$ and $B$ (in contact with $C$) can still be stationary states or, put differently, the final states of $A$ and $B$ coexist when $A$ and $B$ are now brought into contact through a certain contact dynamics. This issue has been addressed in different papers \cite{chatterjee2015zeroth,pradhan2010nonequilibrium,pradhan2011approximate,dickman2016phase}.

The zeroth law of thermodynamics is not expected to hold in full generality for driven systems. However, when chemical potentials at contact can be defined
(see Sec.~\ref{sec:chempot}), the contact dynamics is such that the macroscopic transition rates factorize in terms of the $\phi_{k}$.
One can then attach to each system $k=A,\,B$ the corresponding factor $\phi_{k}(\rho_{k}, \Delta N_{k})$ in the macroscopic transition rate.
One then gets a class of systems that satisfy the zeroth law with respect to each other. Physically speaking, it corresponds to virtually associating one half of the contact to each isolated system in order to build the chemical potential at contact $\mu_{k}^{\mathrm{cont}}$.
The chemical potential $\mu_{k}^{\mathrm{cont}}$ does generically depend on the local contact dynamics and thus cannot be assigned to a purely isolated system but only to the system together with part of the contact. 
In other words, a nonequilibrium chemical potential at contact does not generally obey an equation of state involving only bulk quantities, as recently reported in the context of active particles \cite{guioth2019lack}.
\addreview{Recovering an equation of state would require to tune the contact dynamics with the drive so that it fulfills the condition discussed in Ref.~\cite{chatterjee2015zeroth}.}

\subsection{Relationships between chemical potentials of systems in contact and of isolated systems}
\label{sec:isolated-contact}

\subsubsection{General formula using detailed balance at contact}

We discuss here the relationships between the chemical potentials of systems in contact (see Eq.~\eqref{eq:def:chempot_contact}) and those of isolated systems.

First, one can notice that when macroscopic detailed balance condition \eqref{eq:macro_detailed_balance} as well as the factorization condition \eqref{eq:factorization_micro_transrates} hold, it is sufficient to compute quantities for $\Delta N_{A}= \pm 1$ only since
\begin{align*}
  I'(\rho_{A}|\bar{\rho}) & = \frac{1}{\Delta N_{A}} \ln \frac{\varphi(\rho_{A},- \Delta N_{A})}{\varphi(\rho_{A},\Delta N_{A})} = \ln \frac{\varphi(\rho_{A}, -1)}{\varphi(\rho_{A}, +1)} \\
                          & = \ln \frac{\phi_{A}(\rho_{A},-1)}{\phi_{A}(\rho_{A},+1)} - \ln \frac{\phi_{B}(\rho_{B},-1)}{\phi_{B}(\rho_{B},+1)} 
\end{align*}

Assuming a factorization of the transition rates at the microscopic level as in Eq.~\eqref{eq:factorization_micro_transrates}, the macroscopic transition rates factorize as in \eqref{eq:factorization_transrate} with the factors $\phi_{k}=\lim_{V_{k}\to\infty} \phi_{V_{k},\, k}$ that stem from the finite volume exact expression which reads as
\begin{equation}
  \label{eq:phi_k_factorized}
  \phi_{V_{k},\, k}(\rho_{k}, \pm 1) = \sideset{}{_{c}^{(\pm 1)}} \sum_{\config_{k}^{\prime}} \sideset{}{_{c}^{(0)}} \sum_{\config_{k}} \theta_{k}(\mathcal{C}_{k}', \mathcal{C}_{k}) P_{V_{k}, \, k}(\config_{k}|\rho_{k}) \, .
\end{equation}


\begin{sloppypar}
 In order to get a more insightful expression for $\mu_{k}^{\rm cont} = \ln \left[ \phi_{k}(\rho_{k}, -1)/\phi_{k}(\rho_{k}, +1) \right]$, one should relate $\phi_{k}(\rho_{k}, -1)$ to $\phi_{k}(\rho_{k}, +1)$. As a matter of fact, it is worth considering $\phi_{V_{k},\, k}(-1,\rho_{k}+\tfrac{1}{V_{k}})$ and then take the $V\to\infty$ limit. Using Eq.~\eqref{eq:phi_k_factorized} and the microscopic detailed balance relation in terms of the transition rate factors $\theta_{k}$,
\end{sloppypar}
\begin{equation}
  \label{eq:micro_DB_contact_k}
  \theta_{k}(\mathcal{C}_{k}',\mathcal{C}_{k}) = e^{-\beta(H_{k}(\mathcal{C}_{k}')-H_{k}(\mathcal{C}_{k}))} \theta_{k}(\mathcal{C}_{k},\mathcal{C}_{k}') \, ,
\end{equation}
leads to
\begin{align}
  \label{eq:computation_mucont_factorized_2}
  & \phi_{V_{k},\, k}(\rho_{k}+\tfrac{1}{V_{k}}, -1)  \\
  & \hspace{1em} = \sideset{}{_{c}^{(+1)}} \sum_{\config_{k}^{\prime}} \sideset{}{_{c}^{(0)}} \sum_{\config_{k}} \theta_{k}(\config_{k}^{\prime}, \config_{k}) e^{\beta\left(H_{k}(\config_{k}^{\prime}) - H_{k}(\config_{k}) \right)} \notag \\
  & \hspace{2.1em} \times \frac{P_{V_{k}, \, k}\left(\config_{k}^{\prime} \middle |\rho_{k} + \tfrac{1}{V_{k}}\right)}{P_{V_{k}, \, k}\left(\config_{k}\middle |\rho_{k} \right)} \times P_{V_{k}, \, k}\left(\config_{k}\middle |\rho_{k} \right) \, . \notag
\end{align}

To proceed further, one should insert in Eq.~(\ref{eq:computation_mucont_factorized_2}) explicit expressions of the stationary probability distributions of system $k=A,\, B$ before taking the infinite volume limit. 
It is thus necessary to specify the reference isolated states of systems A and B. We discuss below two cases, on the one hand an equilibrium reference state, and on the other hand the isolated driven system as the reference state.


\subsubsection{Relation between $\mu^{\rm cont}$ and $\mu^{\, \rm eq}$}
\label{sec:link_mucont_mueq}

If one takes the equilibrium state as the reference, $P_{V_{k}, \, k}(\config_{k})$ can be obtained from a perturbative expansion with respect to the equilibrium distribution. This idea was first the one of McLennan \cite{mclennan1959statistical,zubarev1974nonequilibrium} who computed corrections due to the driving force up to first order. Based on this idea to compute perturbatively the nonequilibrium stationary distribution, extensive developments, based on dynamical fluctuations studies, have been performed recently \cite{colangeli2011meaningful,maes2010rigorous,maes2007and,komatsu2008expression,komatsu2009representation,komatsu2008steady,komatsu2010stationary,komatsu2015exact}. One can generally write
\begin{equation}
  \label{eq:McLennan_ansatz}
  P_{V_{k},\, k}(\config_{k}) = \frac{1}{Z_{k}^{\mathrm{eq}}} e^{-\beta H_{k}(\config_{k}) + \Upsilon_{k}^{\mathrm{eq}}(\config_{k})} \, ,
\end{equation}
which defines the supplemental term $\Upsilon_{k}^{\mathrm{eq}}(\config_{k})$ that accounts for the nonequilibrium correction to the Gibbs-Maxwell-Boltzmann equilibrium probability distribution. Introducing this ansatz into Eq.~\eqref{eq:computation_mucont_factorized_2} leads, in the thermodynamic limit, to
\begin{align}
  \label{eq:mucont_correction_wrt_equilibrium_thermolimit}
  & \phi_{k}(\rho_{k},-1) = e^{\mu_{k}^{\mathrm{eq}}(\rho_{k})}  \\
  & \qquad  \times \lim_{V_{k}\to\infty} \!\! \sideset{}{_{c}^{(+1)}} \sum_{\config_{k}^{\prime}} \!\! \sideset{}{_{c}^{(0)}} \sum_{\config_{k}} 
    \!\! \theta_{k}(\config_{k}^{\prime}, \config_{k}) e^{\Upsilon_{k}^{\mathrm{eq}}(\config_{k}^{\prime}) - \Upsilon_{k}^{\mathrm{eq}}(\config_{k}) } \notag \\
  & \hspace{11em} \times P_{V_{k}, \, k}(\config_{k} | \rho_{k}) . \notag
\end{align}
where $\mu_{k}^{\mathrm{eq}} = -\lim_{V_{k}\to\infty} \tfrac{1}{V_{k}} \mathrm{d}\ln Z_{k}^{\mathrm{eq}}/\mathrm{d}\rho_{k}$.

One then notices that $\phi_{k}(\rho_{k}, -1)$ can be related to $\phi_{k}(\rho_{k}, +1)$ through a biased transition rate factor. Indeed
\begin{equation}
  \label{eq:phim1_link_with_bias_transition_rate}
  \phi_{k}(\rho_{k} , -1) = e^{\mu_{k}^{\mathrm{eq}}(\rho_{k})} \phi_{k, \, \Delta \Upsilon_{k}^{\mathrm{eq}} }(\rho_{k}, +1) \, ,
\end{equation}
with $\phi_{k , \, \Delta \Upsilon_{k}^{\mathrm{eq}}}(\rho_{A}, +1)$ the analog of $\phi_{k}(\rho_{A}, +1)$ where $\theta_{k}(\config_{k}^{\prime}, \config_{k})$ has been biased by $\Delta \Upsilon_{k}^{\mathrm{eq}} = \Upsilon_{k}^{\mathrm{eq}}(\config_{k}') - \Upsilon_{k}^{\mathrm{eq}}(\config_{k})$, leading to $\theta_{k}(\config_{k}^{\prime},\config_{k}) e^{\Upsilon_{k}^{\mathrm{eq}}(\config_{k}^{\prime}) - \Upsilon_{k}^{\mathrm{eq}}(\config_{k})}$.

Eventually, according to Eqs.~\eqref{eq:def:chempot_contact} and \eqref{eq:phim1_link_with_bias_transition_rate}, the chemical potential at contact $\mu_{k}^{\mathrm{cont}}$ reads as
\begin{equation}
  \label{eq:mucont_correction_wrt_equilibrium_final}
  \mu_{k}^{\mathrm{cont}}(\rho_{k}) = \mu_{k}^{\mathrm{eq}}(\rho_{k}) + \ln \frac{\phi_{k, \, \Delta \Upsilon_{k}^{\mathrm{eq}}} (\rho_{k}, +1)}{\phi_{k}(\rho_{k}, +1)} \; ,
\end{equation}
where $\ln[ \phi_{k , \, \Delta \Upsilon_{k}^{\mathrm{eq}}}(\rho_{k}, +1)/\phi_{k}(\rho_{k},+1)]$ can be interpreted as an excess chemical potential with respect to the equilibrium one.

One recovers $\mu^{\mathrm{cont}}=\mu^{\mathrm{eq}}$ when the excess nonequilibrium term $\Upsilon_{k}^{\mathrm{eq}}(\config_{k})$ vanishes, meaning that the stationary distribution of the nonequilibrium isolated systems is the same as the equilibrium one. Even though a drive dependence of the stationary distribution is generically expected \cite{maes2009nonequilibrium,colangeli2011meaningful}, one can nevertheless find nonequilibrium models whose stationary solution is not affected by the drive (and is thus equal to the equilibrium one). This is, for instance, the case for the asymmetric simple exclusion process (ASEP) on a ring in one dimension \cite{derrida1998asep}, or for the zero range process \cite{evans2005nonequilibrium}. For this specific subclass of nonequilibrium systems, no shift in stationary densities is expected to be observed when the drives are switched on.
Apart from this small subclass, a shift in densities is generically expected when the drives are switched on.

\subsubsection{Relation between $\mu^{\rm cont}$ and $\mu^{\rm iso}$}
\label{sec:sub:relation_mucont_muiso}
Rather than taking equilibrium as the reference situation, one can also consider the out-of-equilibrium state on its own. Indeed, even if a general procedure to define a nonequilibrium free energy is not yet established, one can sometimes, but rarely, directly compute the nonequilibrium stationary distribution which brings directly an ``out-of-equilibrium partition function'' different from the equilibrium one. Some examples are the zero range process and its extensions \cite{evans2005nonequilibrium,Levine2005,evans2004factorized,evans2006factorized,zia2004construction}, the simple exclusion processes \cite{derrida2007non,derrida1998asep}, etc.

To our knowledge, it does not exist any consensus on a general definition of a genuine nonequilibrium partition function $Z_{k}(\rho)$ for any general system. As already pointed out, however, when correlations are \emph{short ranged},
such a nonequilibrium partition function can be computed by cutting the isolated system into a small, local, macroscopic part, the rest acting as a reservoir. In this case, the stationary probability distribution of the configurations $\config^{\ell}$ of this local part is given by
\begin{equation}
  \label{eq:local_distribution_muiso}
  P_{k}^{\ell}(\config_{k}^{\ell}) = F_{k}(\config_{k}^{\ell}) e^{\mu_{k}^{\mathrm{iso}}(\rho_{k}) \mathcal{N}(\config_{k}^{\ell})} \, ,
\end{equation}
with $F_{k}$ the nonequilibrium weight. The chemical potential $\mu_{k}^{\mathrm{iso}}$ defined in that respect thus fixes the average number of particles inside the isolated system $k$.

In any case, we postulate in the following, without proof, that the stationary distribution $P_{k}$ of the whole system $k$ can be written as
\begin{equation}
  \label{eq:def:out_of_eq_distribution_wout_ref_equilibrium}
  P_{V_{k},\, k}(\config_{k}| \rho_{k}) = \frac{1}{Z_{k}^{\mathrm{neq}}(\rho_{k})}e^{-\beta H_{k}(\config_{k}) + \Upsilon_{k}^{\mathrm{neq}}(\config_{k})}  \, ,
\end{equation}
where $Z_{k}^{\mathrm{neq}}(\rho_{k})$ refers to a nonequilibrium partition function of system $k$, different from the equilibrium one. As observed in a simple but nontrivial nonequilibrium mass transport model \cite{guioth2017mass}, a potential nonequilibrium partition function could be defined such that $\partial \ln Z_{k}^{\mathrm{neq}}/\partial f$ is equal to $\beta$ times the average current, as one would expect for a nonequilibrium generalization of the equilibrium free energy (see \cite{sasa2006steady} for a very detailed discussion on the phenomenological definition of a genuine nonequilibrium free energy). One notices that this expression of the stationary probability distribution can then be obtained from the perturbative expansion \eqref{eq:McLennan_ansatz} simply by introducing the term $e^{-\beta \sigma}$ in \eqref{eq:def:out_of_eq_distribution_wout_ref_equilibrium} where $\sigma$ refers (up to a multiplicative constant length that one takes to be equal to the lattice spacing) to the \addeb{dissipated work} $f\cdot{}J$ when $f$, the nonconservative force, is homogeneous along the system and $J$ is the average current. In this case, $\Upsilon_{k}^{\mathrm{neq}} = \Upsilon_{k}^{\mathrm{eq}} + \beta\sigma$. 
Assuming that the nonequilibrium partition function $Z_{k}^{\mathrm{neq}}(\rho_{k})$ obeys a large-deviations principle with respect to $\rho_{k}$ at the thermodynamic limit, the same calculation as in Sec.~\ref{sec:link_mucont_mueq} leads to
\begin{align}
  \label{eq:phim1_correction_wrt_non-equilibrium}
  &\phi_{k}(\rho_{k}, -1) = e^{\mu_{k}^{\mathrm{iso}}(\rho_{k})} \\
  & \qquad \times \lim_{V_{k}\to\infty} \sideset{}{_{c}^{(+1)}} \sum_{\config_{k}^{\prime}} \!\! \sideset{}{_{c}^{(0)}} \sum_{\config_{k}}  \!
    \theta(\config_{k}^{\prime}, \config_{k}) e^{\Upsilon_{k}^{\mathrm{neq}}(\config_{k}^{\prime}) - \Upsilon_{k}^{\mathrm{neq}}(\config_{k}) } \notag \\
  & \hspace{11.3em}\times P_{V_{k},\, k}(\config_{k} | \rho_{k}) , \notag 
\end{align}
which in turn gives
\begin{equation}
  \label{eq:mucont_correction_wrt_non-equilibrium}
  \mu_{k}^{\mathrm{cont}}(\rho_{k}) = \mu_{k}^{\mathrm{iso}}(\rho_{k}) + \ln \frac{\phi_{k , \, \Delta \Upsilon_{k}^{\mathrm{neq}}}(\rho_{k}, +1)}{\phi_{k}(\rho_{k},+1)} \; .
\end{equation}
This expression is almost identical to the previous one \eqref{eq:mucont_correction_wrt_equilibrium_final} for which the reference situation was the equilibrium. Different normalizations of the stationary probability distribution can thus lead to different chemical potentials in excess with respect to the chosen reference configuration. 


\subsubsection{Contact dependence of the excess chemical potential}

In each case, one sees that the chemical potential at contact $\mu_{k}^{\mathrm{cont}}(\rho_{k})$ is equal to a chemical potential related to the isolated system (either the equilibrium one or the stationary nonequilibrium one) and an \emph{excess} chemical potential which generically reads as
\begin{equation}
  \label{eq:excess_chemical_potential}
  \eta_{k}(\rho_{k}) = \ln \frac{\phi_{k , \, \Upsilon_{k}}(\rho_{k}, +1)}{\phi_{k}(\rho_{k}, +1)} \, .
\end{equation}
The excess chemical potential $\eta_{k}$ clearly depends on $\Delta \Upsilon_{k}$ that has to be different from $0$ to get $\eta_{k}$ nonvanishing.

We should however recognize that one cannot avoid any \emph{generic} dependence on $\theta_{k}(\config_{k}^{\prime}, \config_{k})$ in the expression of $\eta_{k}$ as long as the steady-state measure of isolated systems is affected by the drives. That is to say, the details of the contact, which involves mobility (symmetric part) and force (antisymmetric part) of the contact, do contribute to the chemical potential $\mu_{k}^{\mathrm{cont}}$. One cannot exclude as of now that the chemical potential at contact $\mu_{k}^{\mathrm{cont}}$ may depend on the details of the contact dynamics, beyond the specificities of stationary probability distributions of the isolated systems.

We thus confirm and significantly extend preliminary results obtained by Sasa, Hayashi and Tasaki \cite{hayashi2003thermo,sasa2006steady} on the KLS model, although the latter results were not recognized as resulting from a large-deviations analysis.

\subsection{Driven contact dynamics}
\label{sec:extra-work_tilt}

\subsubsection{Driven systems with a drive-dependent contact dynamics}

Until now, we have considered a contact orthogonal to the nonconservative driving forces, leading to transition rates at contact independent of the driving forces of systems $A$ and $B$ and verifying detailed balance with respect to the equilibrium distributions. One can nevertheless wonder what happens when microscopic transition rates at contact do depend on the driving forces or when there is an extra work performed at contact. In that respect, if one assumes that transition rates at contact obey a local detailed balance with extra work in addition to the local energy difference present at equilibrium, the same reasoning still applies.

We assume that the microscopic transition rate at contact, $T_{c}(\config^{\prime}|\config)$ obeys local detailed balance with additional work $w^{\mathrm{cont}}(\config,\config^{\prime})$ that can depend on the driving forces themselves. We consider also a possible extra influence of the driving forces on the symmetric part of the transition rates $a(\config,\config^{\prime})$. One then obtains
\begin{equation}
  \label{eq:extra_dependences_transrate_contact}
  T_{c}(\config^{\prime}|\config) = a^{\mathrm{neq}}(\config,\config^{\prime}) \, e^{-\tfrac{\beta}{2}\left(H(\config^{\prime}) - H(\config) - w^{\mathrm{cont}}(\config,\config^{\prime})\right)} \, .
\end{equation}
Assuming that both the factorization condition \eqref{eq:factorization_micro_transrates} and macroscopic detailed balance hold, the same calculation presented in last subsection \ref{sec:isolated-contact}, valid at the vanishing exchange rate limit, leads to
\begin{align}
  \label{eq:ratio_extra_macro_transrate}
  \mu_{k}^{\mathrm{cont}}(\rho_{k}) & \equiv \ln \frac{\phi_{k}(\rho_{k}, -1)}{\phi_{k}(\rho_{k},+1)} \\
  &  = \mu_{k}(\rho_{k}) + \ln \frac{\phi_{k , \, \Delta \Upsilon_{k} + w_{k}^{\mathrm{cont}} }(\rho_{k},+1)}{\phi_{k}(\rho_{k}, +1)} . \notag
\end{align}
Not surprisingly, one can see that the excess chemical potential due to the additional work $w_{k}^{\mathrm{cont}}$ (resulting from the splitting of $w^{\mathrm{cont}}$ into two terms ($k=A,B$) from the factorization condition~\eqref{eq:factorization_micro_transrates} of the transition rate~\eqref{eq:extra_dependences_transrate_contact}) is added to the out-of-equilibrium term $\Delta \Upsilon_{k}$ (a term breaking microscopic detailed balance by making $P(\config_{k}|\bar{\rho_{k}})$ different from the equilibrium distribution).

It may happen that the presence of the additional work $w^{\mathrm{cont}}$ needed to realize the transition $\config \to \config'$ could break the factorization property. If this is so, one has to come back to the global expression of the derivative of the large-deviations function which would thus be nonadditive. For situations when this extra work is only exerted through the contact by external agents and thus is not a function of energies or applied work in each systems -- in short, when $w^{\mathrm{cont}}$ depends neither on $A$ nor on $B$ --, the splitting into two contributions to get the factorization property could still be made, even though perhaps in a quite arbitrary way, by considering the work needed to move particles from $A$ to $B$ and conversely to move a particle from $B$ to $A$.

\subsubsection{Equilibrium systems with an active contact}
Even at equilibrium, the case where the microscopic transition rates involve an extra work is quite interesting. In particular, this situation happens in biological systems and more specifically in cells. Indeed, let us consider two compartments separated by a membrane punctuated by channels that allow the transfer of particles (ions here) from one side to another. Two types of channels have been observed \cite{siwy2002ionpump,gadsby2009ionspumps}. The first type concerns passive channel (called \emph{ion channel}) that let the ions follow the local electrochemical potential gradient (which basically embeds, if one neglects interactions between ions, simple diffusion and possible complex electric potential across the membrane). But, in several situations it is observed that the transfer of particles is not passive and does not follow the electrochemical potential gradient. At a microscopic level, this is due to the fact that the channel is active and consumes metabolic energy to transport ions. These are usually referred to as \emph{ion transporters} or \emph{ion pumps}. Thus, this active transport involves an extra work $w^{\mathrm{cont}}$ which is localized in the channel. If the frequency of exchange is very small and if one takes for reference an equilibrium situation where in each system $k$ the stationary probability distribution reads as $P_{k}^{\mathrm{eq}}(\config_{k}|\rho_{k}) = e^{-\beta H_{k}(\config_{k})}/Z_{k}^{\mathrm{eq}}$, one obtains
\begin{equation}
  \label{eq:mu_cont_eq_nonconservative_work_contact}
  \mu_{k}^{\mathrm{cont}}(\rho_{k}) = \mu_{k}^{\mathrm{eq}}(\rho_{k}) + \ln \frac{\phi_{k, \, w_{k}^{\mathrm{cont}}}(\rho_{k}, +1)}{\phi_{k}(\rho_{k},+1)} \, ,
\end{equation}
if both macroscopic detailed balance and factorization conditions hold. In the case where the active transport is switched off, $w_{k}^{\mathrm{cont}}$ vanishes, and one recovers the passive equilibrium potentials $\mu_{k}^{\mathrm{eq}}(\rho_{k})$. 

\subsection{Chemical potential and external potential}

An operational way to define and measure a nonequilibrium chemical potential has been put forward in \cite{sasa2006steady}. The idea is to apply a potential energy difference $\Delta U=U_A-U_B$ between the two driven systems $A$ and $B$ in contact.
According to Sasa and Tasaki \cite{sasa2006steady}, the nonequilibrium chemical potentials are the functions $\mu_A(\rho_A)$ and $\mu_B(\rho_B)$ that satisfy
\begin{equation} \label{eq:ST:DeltaU}
\mu_{A}(\rho_A) + U_{A} = \mu_{B}(\rho_B) + U_{B} \, ,
\end{equation}
where $\rho_A$ and $\rho_B$ are the steady-state densities measured in systems $A$ and $B$ when the potential energy difference $\Delta U$ is switched on.
Note that the functions $\mu_{A}(\rho_A)$ and $\mu_{B}(\rho_B)$ do not depend explicitly on the applied potential difference $\Delta U$.

Our present general framework allows us to determine the validity conditions of the phenomenologically postulated Eq.~(\ref{eq:ST:DeltaU}).
It is convenient to use the decomposition of the coarse-grained transition rate $\varphi_{\Delta U}(\rho,\Delta N)$ in terms of thermodynamic force and mobility (see Eq.~\eqref{eq:decomposition_force_act_transrate})
\begin{equation}
\varphi_{\Delta U}(\rho,\Delta N) = a_{\Delta U}(\rho,\Delta N) e^{\frac{1}{2}F_{\Delta U}(\rho,\Delta N)} \, .
\end{equation}

It turns out that with our definition (\ref{eq:def:chempot_contact}) of the chemical potential,
Eq.~(\ref{eq:ST:DeltaU}) is valid only under the assumptions that macroscopic detailed balance holds and that the coarse-grained transition rate $\varphi_{\Delta U}(\rho,\Delta N)$ has a specific dependence on $\Delta U$, namely
(i) the symmetric part $a_{\Delta U}(\rho,\Delta N)$ is independent of $\Delta U$
and (ii) the antisymmetric part $F_{\Delta U}(\rho,\Delta N)$ is linear with $\Delta U$, i.e., $F_{\Delta U}=F + \Delta U \Delta N$.
For other forms of the transitions rates that do not satisfy the above conditions (i) and (ii), Eq.~(\ref{eq:ST:DeltaU}) does not hold with our definition of the chemical potential.



\section{Explicit examples of lattice gas models in contact}
\label{sec:lattice_models}

We now apply the general framework to lattice models in contact, on the example of the model recently introduced in \cite{guioth2017mass}. 
This model has been chosen over more standard lattice models \cite{spitzer1970,liggett2012interacting,evans2005nonequilibrium,zia2010twenty,katz1984nonequilibrium} because its steady-state distribution can be determined exactly and it depends on the nonequilibrium driving force, a generic property according to, for instance, the McLennan expansion \cite{mclennan1959statistical}.
In contrast, standard models like the zero range process (ZRP) \cite{spitzer1970,liggett2012interacting,evans2005nonequilibrium} or the asymmetric simple exclusion process (ASEP) \cite{spitzer1970,derrida1998asep} with periodic boundary conditions have a steady-state distribution that is independent of the drive.
Other models, like the KLS model \cite{katz1984nonequilibrium,zia2010twenty}, are expected to have a steady-state distribution that depends on the drive, but this distribution is not known exactly.

\subsection{Exactly solvable driven lattice gas model}
\label{sec:MTM}

\subsubsection{Definition and steady-state distribution}

As for the ZRP, one considers a one-dimensional lattice $\Lambda$ of $|\Lambda|$ sites. The number of sites is assumed to be even and we write $|\Lambda|=2L$ with $L$ an integer. Each site $i$ is occupied by $n_{i}\geqslant 0$ particles that cannot exceed a maximum number of $n_{\mathrm{max}}$ particles per site, that may be infinite. The dynamics of this model is inspired by an equilibrium KCM (kinetically constrained models) model \cite{PhysRevLett.95.015702} as well as by the ZRP \cite{evans2005nonequilibrium}. Contrary to most of the standard mass transport models, the dynamics is \emph{synchronous} and involves two partitions of the lattice, namely, $\mathcal{P}_{1}=\{(2k,2k+1)\}_{k\in[ 0, L]}$ and $\mathcal{P}_{2}=\{(2k+1,2k+2)\}_{k\in [ 0, L ]}$ which gather alternate adjacent pairs of sites. A partition is selected randomly between $\mathcal{P}_{1}$ and $\mathcal{P}_{2}$ with equal probability. Once a partition $\mathcal{P}_{j}$ has been chosen, all links belonging to $\mathcal{P}_{j}$ are updated in parallel and independently, and a link $(i,i+1)$ is updated according to the following transition probability,
\begin{align}
  \label{eq:def:transrate_MTM_model}
  &T(n_{i+1}^{\prime},n_{i}^{\prime}|n_{i+1},n_{i}) = K(d_{i}^{\prime} | \bar{n}_{i}) \\
  & \hspace{1em} = \frac{1}{Q(\bar{n}_{i})}\,
\exp\left\{ -\left[\varepsilon\left(\bar{n}_{i}+\tfrac{d_{i}^{\prime}}{2}\right)+\varepsilon\left(\bar{n}_{i}-\tfrac{d_{i}^{\prime}}{2}\right)\right] + \tfrac{f}{2} d_{i}^{\prime} \right\} . \notag
\end{align}
with $n_{i+1}^{\prime}+n_{i}^{\prime}= n_{i+1}+n_{i}$ since particle number is conserved ($T=0$ otherwise). Notations $d_{i}^{\prime} = n_{i+1}^{\prime}-n_{i}^{\prime}$ and $\bar{n}_{i}=(n_{i}+n_{i+1})/2=(n_{i}^{\prime}+n_{i+1}^{\prime})/2$ have been introduced.
The normalization factor $Q(\bar{n}_{i})$ is such that $\sum_{n_{1}^{\prime},n_{2}^{\prime}} T(n_{2}^{\prime},n_{1}^{\prime}|n_{1},n_{2}) = 1$. We emphasize that the net transfer of particles $\Delta n_{i}=(n_{i}^{\prime}-n_{i})=-(n_{i+1}^{\prime}-n_{i+1})$ from site $i$ to site $i+1$ is given by $\Delta n_{i} = (d_{i}^{\prime}-d_{i})/2$ with $d_{i}=n_{i+1}-n_{i}$.
The probability to choose a \comebb{difference of particle numbers} $d_{i}^{\prime}$ between sites $i$ and $i+1$ is independent of $d_{i}$ which means that the probability to transfer $\Delta n_{i}$ particles does not depend on the initial difference \comebb{of particle numbers} of the two sites, as one might expect intuitively for a mass transport model. In the absence of $f$, the uniform energy $\varepsilon(n)$ attached to each site tends to homogenize the density over the link. The parameter $f$ can be interpreted as a driving force since it pushes particles toward the site $i+1$ whatever the initial configuration. In terms of local detailed balance, one has:
\begin{equation}
  \label{eq:ldb_MTM}
  \ln\frac{T(n_{i+1}^{\prime},n_{i}^{\prime}|n_{i+1},n_{i})}{T(n_{i+1},n_{i}|n_{i+1}^{\prime},n_{i}^{\prime})}=-\Delta \varepsilon_{i+1}-\Delta \varepsilon_{i} + f\Delta n_{i}
\end{equation}
with $\Delta \varepsilon_{i}=\varepsilon(n_{i}^{\prime})-\varepsilon(n_{i})$. Hence, Eq.~\eqref{eq:ldb_MTM} confirms the interpretation of $f$ as a driving force since $f\Delta n_{i}$ can be interpreted as the work needed to move a number of particles $\Delta n_{i}$ from site $i$ to site $i+1$. As for the alternation of the partition choice, one can imagine that this is produced by an oscillating confining potential of two sites period. In the presence of a driving force $f$, the oscillating potential would look more like a sawtooth potential of slope $-f$.

The stationary probability density function can be exactly computed (see \cite{guioth2017mass} for the continuous mass version of this model) and reads as
\begin{align}
  \label{eq:stationary_probability_MTM}
  & P\left( \{ n_{i} \}_{i\in\Lambda} \right) \\
  & \hspace{1em} = \frac{2}{Z(|\Lambda|,N)} \exp\left( \sum_{i\in\Lambda} \varepsilon(n_{i})\right) \cosh\left( \sum_{i\in \Lambda} (-1)^{i}fn_{i} \right) . \notag
\end{align}
One thus observes that the stochastic oscillating forcing produces long-range static correlations that can be anticipated from the presence of the hyperbolic cosine factor in \eqref{eq:stationary_probability_MTM} (see \cite{guioth2017mass} for detailed calculations of the static two-points correlation function).

\addeb{Note that in \cite{guioth2017mass}, the dynamics of the model was defined in discrete time, which is a natural framework to deal with synchronous dynamics. Here, however, we consider a continuous time synchronous dynamics, meaning that the synchronous updates of lattice partitions occur at random continuous times. Note that the stationary probability distribution is the same with discrete and with continuous time. Although a continuous time dynamics might seem artificial in the context of a synchronous update, its use allows for an easier implementation of a contact dynamics between two systems, as discussed below.
}

\subsection{Two driven lattice gas models in contact}
\label{sec:def_contact_MTM}

We now move to the study of the contact between two different systems, say $A$ and $B$. As for the ZRP case, one needs to define precisely the dynamics at contact since only isolated systems have been defined so far. We want to connect both systems to each other with at least one link, say $i_{A}\in\Lambda_{A}$ and $j_{B}\in\Lambda_{B}$. But contrary to the bulk dynamics for which all links are updated in parallel, the contact dynamics is assumed not to be synchronous with respect to the bulk. 
An exchange between both systems is thus selected at a rate very small compared to the bulk one of each system. The dynamics at contact needs to satisfy local detailed balance in the absence of drive since there is \emph{a priori} no reason that the transition rates change when systems are driven out-of-equilibrium \emph{orthogonally} to the contact. Since an energy $\varepsilon(n_{i})$ is attached to each site filled by $n_{i}$ particles, we set $T_{c}$, the transition rate at contact, such that it satisfies the local detailed balance that reads as
\begin{align}
  \label{eq:local_detailed_balance_MTM}
  & \frac{T_{\mathrm{c}}(n_{i_{A}}',n_{j_{B}}'|n_{i_{A}},n_{j_{B}})}{T_{\mathrm{c}}(n_{i_{A}}, n_{j_{B}}|n_{i_{A}}', n_{j_{B}}')} \\
  & \hspace{2em} = e^{-\left[\varepsilon_{A}(n_{i_{A}}') - \varepsilon_{A}(n_{i_{A}})\right]}  e^{-\left[ \varepsilon_{B}(n_{j_{B}}') - \varepsilon_{B}(n_{j_{B}}) \right]}, \notag
\end{align}
if particle number conservation $n_{i_{A}}'+n_{j_{B}}' = n_{i_{A}} + n_{j_{B}}$ holds, and $T_{c}=0$ otherwise. From here on, we consider different dynamics that will differ by a different choice of the mobility parameter \cite{maes2007and,maes2008canonical} --we recall that the mobility refers to the parameter $a$ in the decomposition $T_{c}(\config'|\config) = a(\config,\config')\, \exp[\tfrac{1}{2}
F(\config,\config')]$ according to which $a(\config,\config')=a(\config',\config)$ and $F(\config,\config')=-F(\config',\config)$.

\subsubsection{Natural dynamics}

We first consider the case when the transition rate at contact is similar to the dynamics in the bulk, i.e., the transition rate depends on the final configuration:
  \begin{equation}
    \label{eq:NatDyn_rule_contact_MTM}
    T_{\mathrm{c}}(n_{i_{A}}',n_{j_{B}}'|n_{i_{A}},n_{j_{B}}) \propto e^{-\varepsilon_{A}(n_{i_{A}}')} e^{-\varepsilon_{B}(n_{j_{B}}')} .
  \end{equation}
The symbol $\propto$ means here that the transition rates are equal to the right-hand side up to a constant factor that sets the typical time scale associated with the transition. In the slow exchange contact limit, the latter factor will be infinitesimally small. 

\subsubsection{Sasa-Tasaki rule}

For the Sasa-Tasaki rule which models a high energy barrier separating both systems, the probability to transfer a particle from $A$ to $B$ (respectively from $B$ to $A$) only depends on the energy to go from the $A$ side (respectively $B$ side) bottom of the barrier to its top. Hence, it reads as
  \begin{align}
    \label{eq:ST_rule_contact_MTM}
    &T_{\mathrm{c}}(n_{i_{A}}',n_{j_{B}}'|n_{i_{A}},n_{j_{B}}) \\
    &\hspace{1em}
      \propto
      \begin{cases}
      \exp\left\{-\left[\varepsilon_{A}(n_{i_{A}}')-\varepsilon_{A}(n_{i_{A}})\right] \right\} \text{ if } n_{i_{A}}' < n_{i_{A}} \\
      \exp\left\{ -\left[\varepsilon_{B}(n_{j_{B}}') -\varepsilon_{B}(n_{j_{B}}) \right] \right\} \text{ if } n_{i_{A}}' > n_{i_{A}}  
    \end{cases}
    \notag
  \end{align}

\subsubsection{Kawasaki or heat bath rule}

The Kawasaki, or heat-bath, rule, is a standard choice of transition rate.
It does not factorize in two terms that respectively depend on $A$ and $B$:
  \begin{align}
    \label{eq:Kawasaki_rule_contact_MTM}
    &T_{\mathrm{c}}(n_{i_{A}}',n_{j_{B}}'|n_{i_{A}},n_{j_{B}}) \\
    & \hspace{2em} \propto \frac{2}{1+e^{ \left[\varepsilon_{A}(n_{i_{A}}')-\varepsilon_{A}(n_{i_{A}})\right] + \left[\varepsilon_{B}(n_{j_{B}}')-\varepsilon_{B}(n_{j_{B}}) \right]} } . \notag
  \end{align}
\addeb{Note that another standard and qualitatively similar transition rate is the Metropolis rule. In what follows, we shall use only the Kawasaki rate for the purpose of illustration, but similar results can be obtained with the Metropolis rule.}

\subsection{Large-deviations function and chemical potentials for single-particle exchange}

Having specified the dynamics, we will compute here the large-deviations function of the density. Under the hypothesis that the exchange of particles between systems is very rare, the coarse-grained transition rate \eqref{eq:coarse_grained_transrate_rescaled_time} reads as
\begin{align}
  \label{eq:coarse_grained_transrate_MTM}
  \varphi(\rho_{A}, \Delta n) & = \hspace{-5pt} \sum_{n_{i_{A}}, n_{j_{B}}} \hspace{-5pt} T( n_{i_{A}}+\Delta n, n_{j_{B}}-\Delta n | n_{i_{A}}, n_{j_{B}} ) \\
  & \hspace{1em} \times P(n_{i_{A}}|\rho_{A}) P(n_{j_{B}}|\rho_{B}) . \notag 
\end{align}

When only one particle can be exchanged, the macroscopic detailed balance \eqref{eq:macro_detailed_balance} always holds. To illustrate the dependence of the large-deviations function, and thus the chemical potentials when defined, with respect to the dynamics at contact, we compute the latter for the three contact dynamics presented in Sec.~\ref{sec:def_contact_MTM}.

\subsubsection{Natural dynamics and the Sasa-Tasaki rule}

We start by considering the natural dynamics \eqref{eq:NatDyn_rule_contact_MTM} and the Sasa-Tasaki rule \eqref{eq:ST_rule_contact_MTM} as the dynamics of the contact.
Since these microscopic dynamics are factorized, the coarse-grained transition rates also take a factorized form,
\begin{equation}
\varphi(\rho_{A},\Delta N_A) = \phi_A(\rho_{A},\Delta N_A) \phi_B(\rho_{B},\Delta N_B)
\end{equation}
with $\Delta N_B = -\Delta N_A = \pm 1$.

The explicit expressions of the factors $\phi_{A}(\rho_{A}, \Delta N_{A})$ and $\phi_{B}(\rho_{B}, \Delta N_{B})$ for each dynamics are given in Appendix~\ref{sec:appB}.
Transition rates being factorized, one can associate with each systems chemical potentials that read as, according to \eqref{eq:def:chempot_contact},

\begin{equation}
  \label{eq:chem_pot_ND_ST_MTM}
  \mu_{k}^{\rm cont}(\rho_{k}) = \mu_{k}^{\rm iso}(\rho_{k}) + \eta_{k}(\rho_{k})
\end{equation}
with $\mu_{k}^{\rm iso}$ given by Eq.~\eqref{eq:mu_iso_implicit_def} for every dynamics and $\eta_{k}$ reading as
\begin{align}
  \label{eq:excess_chem_pot_ND_ST_MTM}
  & \eta_{k}^{\rm (ND)}(\rho_{k})  \\
  & \; = \ln \frac{ \sum_{n_{k}=0}^{n_{\mathrm{max}}^{k}-1} e^{-\left[\varepsilon_{k}(n_{k})+\varepsilon_{k}(n_{k}+1)\right] + \mu_{k}^{\mathrm{iso}}n_{k}} e^{\upsilon[\mu_{k}^{\mathrm{iso}}, f_{k}](n+1) }}{\sum_{n_{k}=0}^{n_{\mathrm{max}}^{k}-1} e^{-\left[\varepsilon_{k}(n_{k})+\varepsilon_{k}(n_{k}+1)\right] + \mu_{k}^{\mathrm{iso}} n_{k} } e^{\upsilon[\mu_{k}^{\mathrm{iso}}, f_{k}](n)} } \, , \notag \\
  & \eta_{k}^{\rm (ST)}(\rho_{k})  =   \ln \frac{\sum_{n_{k}=0}^{n_{\mathrm{max}}^{k}-1}e^{-\varepsilon_{k}(n_{k})+\mu_{k}^{\mathrm{iso}}n_{k}} e^{\upsilon[\mu_{k}^{\mathrm{iso}},f_{k}](n_{k}+1)} }{ \sum_{n_{k}=0}^{n_{\mathrm{max}}^{k}-1}e^{-\varepsilon_{k}(n_{k})+\mu_{k}^{\mathrm{iso}}n_{k}} e^{\upsilon[\mu_{k}^{\mathrm{iso}},f_{k}](n_{k})}} , \notag
\end{align}
with  $\upsilon[\mu_{k}^{\mathrm{iso}}, f_{k}]$ given in Appendix~\ref{sec:appB} [see Eq.~(\ref{def:eq:upsilon:mu:f})].
The expression of the excess chemical potentials, and thus of the chemical potentials of the systems in contact, take different forms for both dynamics, as expected from the generic dependence of the nonequilibrium chemical potentials on the contact dynamics.
The difference between the two contact dynamics will be discussed quantitatively in Sec. \ref{sec:num_sim_examples}.

\subsubsection{Kawasaki rule}

As a last example, we turn to the Kawasaki rule \eqref{eq:Kawasaki_rule_contact_MTM}, for which the microscopic dynamics does not take a factorized form.
The coarse-grained transition rate reads as
\begin{align}
  \label{eq:transrate_Kawasaki_1part_MTM}
&  \varphi(\rho_{A}, +1)   \\
&   \hspace{0.5em} = \sum_{n_{i_{A}}=0}^{n_{\mathrm{max}}^{A}-1} \sum_{n_{j_{B}}=1}^{n_{\mathrm{max}}^{B}} \frac{2 P(n_{i_{A}}|\rho_{A}) P(n_{j_{B}} | \rho_{B})}{1 + e^{\varepsilon_{A}(n_{i_{A}}+1)-\varepsilon_{A}(n_{i_{A}}) + \varepsilon_{B}(n_{j_{B}}-1) - \varepsilon_{B}(n_{j_{B}}) }  } \notag \\
 & \varphi(\rho_{A}, -1)   \notag \\
  & \hspace{0.5em} = \sum_{n_{i_{A}}=1}^{n_{\mathrm{max}}^{A}} \sum_{n_{j_{B}}=0}^{n_{\mathrm{max}}^{B}-1} \frac{2 P(n_{i_{A}}|\rho_{A}) P(n_{j_{B}} | \rho_{B})}{1 + e^{\varepsilon_{A}(n_{i_{A}}-1)-\varepsilon_{A}(n_{i_{A}}) + \varepsilon_{B}(n_{j_{B}}+1) - \varepsilon_{B}(n_{j_{B}}) }  } \notag \, .
\end{align}
Here the coarse-grained transition rates do not factorize, so that the large-deviations function is not additive, implying that a chemical potential cannot be defined. One can nevertheless evaluate the derivative of the large-deviations function \eqref{eq:macro_detailed_balance}, which reads as
\begin{widetext}
  \begin{multline}
\label{eq:Iprime:Kawa:MTM}
  I'(\rho_{A}|\bar{\rho}) = \mu_{A}^{\mathrm{iso}} - \mu_{B}^{\mathrm{iso}}
  + \ln \left[ \sum_{n_{i_{A}}=0}^{n_{\mathrm{max}}^{A}-1} \sum_{n_{j_{B}} = 0}^{n_{\mathrm{max}}^{B}-1} \frac{ 2 e^{ \mu_{A}^{\mathrm{iso}} n_{i_{A}} + \mu_{B}^{\mathrm{iso}} n_{j_{B}} } e^{\upsilon[\mu_{A}^{\mathrm{iso}},f_{A}](n_{i_{A}}+1) + \upsilon[\mu_{B}^{\mathrm{iso}},f_{B}](n_{B})}}{e^{\varepsilon_{A}(n_{i_{A}}+1)+\varepsilon_{B}(n_{j_{B}})} + e^{\varepsilon_{B}(n_{j_{B}}+1)+ \varepsilon_{A}(n_{i_{A}}) }} \right] \\
  - \ln \left[ \sum_{n_{i_{A}}=0}^{n_{\mathrm{max}}^{A}-1} \sum_{n_{j_{B}} = 0}^{n_{\mathrm{max}}^{B}-1} \frac{ 2 e^{ \mu_{A}^{\mathrm{iso}} n_{i_{A}} + \mu_{B}^{\mathrm{iso}} n_{j_{B}} } e^{\upsilon[\mu_{A}^{\mathrm{iso}},f_{A}](n_{i_{A}}) + \upsilon[\mu_{B}^{\mathrm{iso}},f_{B}](n_{B}+1)}}{e^{\varepsilon_{A}(n_{i_{A}}+1)+\varepsilon_{B}(n_{j_{B}})} + e^{\varepsilon_{B}(n_{j_{B}}+1)+ \varepsilon_{A}(n_{i_{A}}) }} \right] \; .
  \end{multline}
\end{widetext}
Equating this derivative to zero still allows for a characterization of the stationary densities of the systems in contact. However, this characterization cannot be written as the equality of chemical potentials depending only on the properties of a given system (even including contact properties).
Rather, equating the expression \eqref{eq:Iprime:Kawa:MTM} of $I'(\rho_{A}|\bar{\rho})$ to zero yields the equality of two functions that both depend on the two densities $\rho_A$ and $\rho_B$.

\subsubsection{Comments on the contact dynamics}

Before concluding this subsection on the evaluation of the chemical potential in the lattice gas model, two comments are in order.
The first one is that when there is at most one particle on each site, i.e.,
$n_{\mathrm{max}}^{k}=1$ for both systems, $P(n_{k}|\rho_{k})= \rho_{k}$ by translation symmetry and is thus independent of the driving force $f$. In this case, one recovers an equilibrium situation and stationary densities are given by the equality of the equilibrium chemical potential $\mu_{k}^{\mathrm{cont}}(\rho_{k}) = \mu_{k}^{\mathrm{eq}}(\rho_{k}) = \ln(\rho_{k}/(1-\rho_{k}))$. 

The second comment concerns situations when the contact between the two systems is extended along several links.
Up to now, the contact was built along a single link involving only two sites. In general, several links may be involved in the contact area. But since the dynamics is asynchronous, only one link can be chosen at a time,
and observing any effect related to the extension of the contact area is not expected. Numerical simulations performed confirm this hypothesis (see below).

\subsection{Numerical simulations and explicit examples}\label{sec:num_sim_examples}


In all cases studied, the excess chemical potentials, or excess large-deviations derivative, is nonzero because of the presence of the nonequilibrium \comebb{factor $e^{\upsilon[\mu^{\mathrm{iso}},f](n)}$ appearing in Eq.~(\ref{eq:excess_chem_pot_ND_ST_MTM})}. But one can wonder what is the magnitude of these correction terms compared to the chemical potential of the isolated systems. In order to address this question, we provide some plots of the chemical potentials at contact in different situations.

\begin{figure}[t]
  \includegraphics[width=0.9\linewidth]{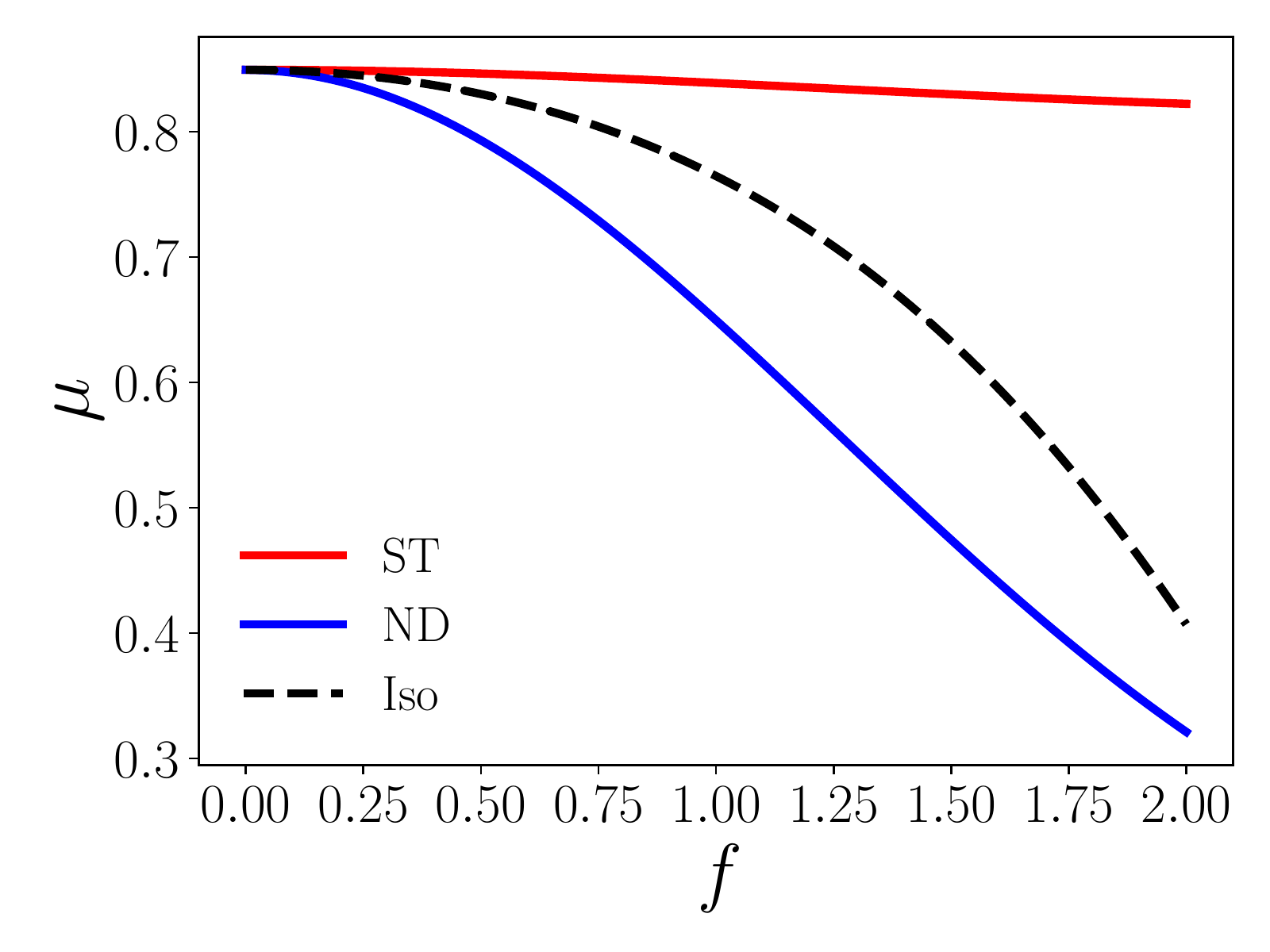}
  \caption{Plots of the chemical potentials associated with one system. The parameters are $n_{\mathrm{max}}=2$, $\rho=0.9$, and an energy parameter $\varepsilon_{0}=1$ (for $\varepsilon(n)=\varepsilon_{0} n$). The chemical potentials $\mu^{\mathrm{iso}}$ (dashed black curve), $\mu^{\rm cont, (ND)}$ (blue curve) and $\mu^{\mathrm{cont, (ST)}}$ (red curve) are plotted as functions of the forcing $f$.}
  \label{fig:chem_pot_plot_MTM}
\end{figure}

We fix the maximum number of particles to $n_{\mathrm{max}}=2$ and we choose a simple linear energy function $\varepsilon(n)=\varepsilon_{0} n$. The first figure (see Fig.~\ref{fig:chem_pot_plot_MTM}) represents the chemical potentials at contact both for the natural and Sasa-Tasaki dynamics as well as the chemical potential associated with the isolated system with respect to the driving force $f$ at a fixed density $\rho=0.9$. For $f\geqslant 0.25$, the three different chemical potentials start to differ significantly and one may thus expect a clear effect of the drive coupled to the specific contact dynamics at play.

\begin{figure}[h]
\includegraphics[width=0.95\linewidth]{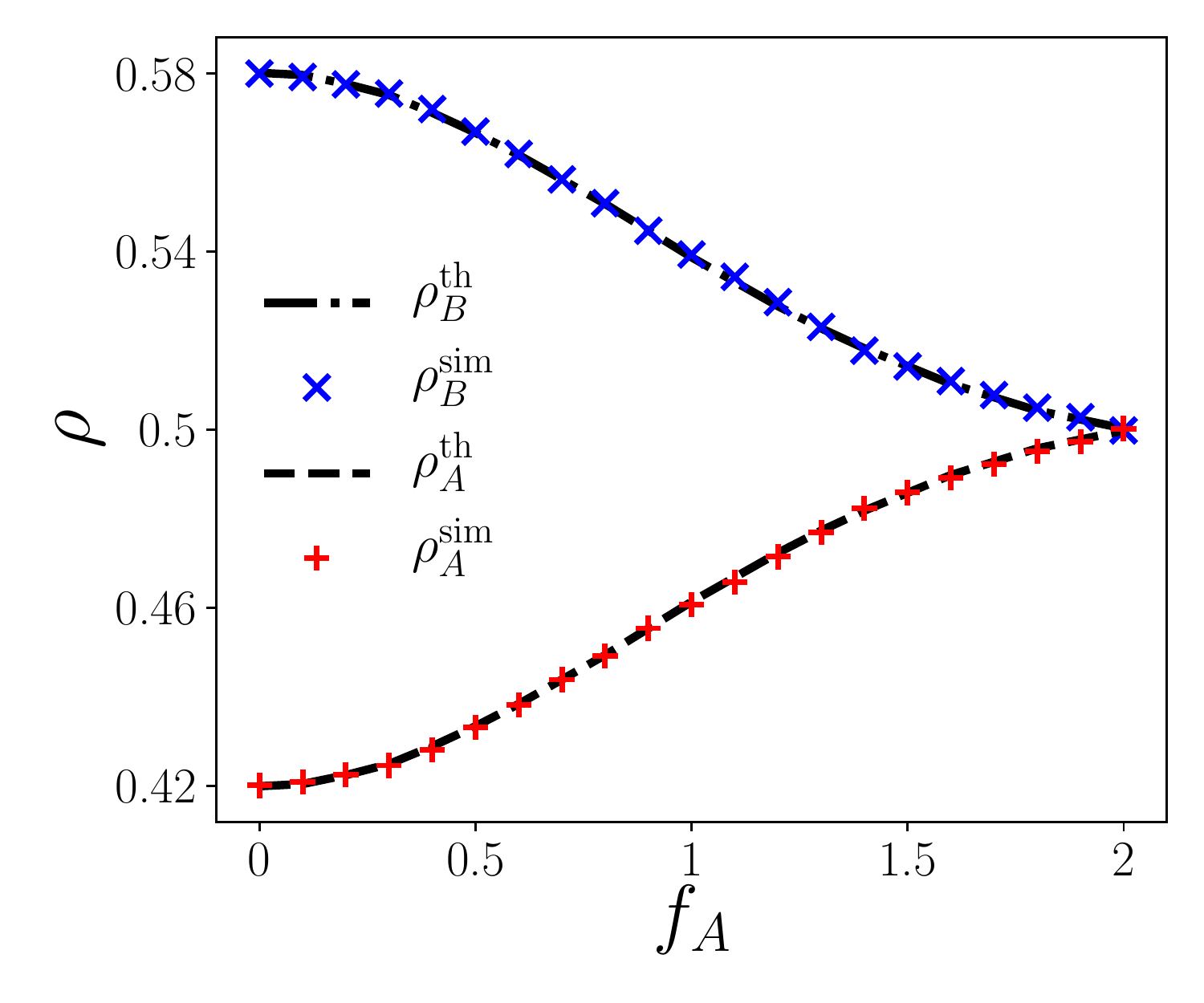}\\
\includegraphics[width=0.95\linewidth]{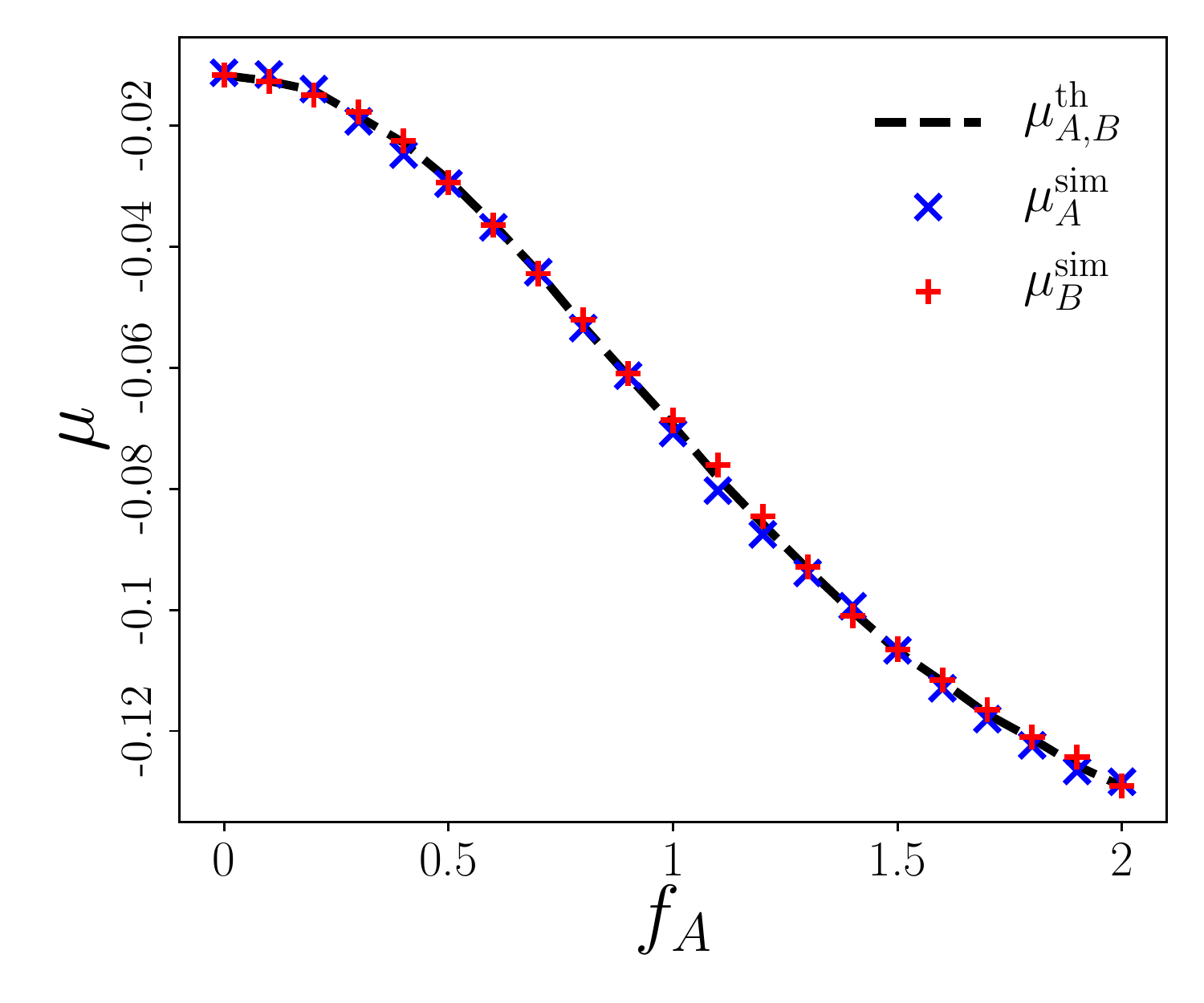}
  \caption{Numerical simulations of two lattice models $A$ and $B$ in contact, with different drives.
Top: Densities $\rho_A$ (red) and $\rho_B$ (blue) versus time.
Bottom: Chemical potentials $\mu_{A}^{\rm cont}$ (red) and $\mu_{B}^{\rm cont}$ (blue).
The dashed lines are the theoretical predictions. Parameters: $|\Lambda_A|=|\Lambda_B|=10000$, $\bar{\rho}=0.5$. The dynamics used is the ``natural dynamics'' (ND).}
\label{fig:numerical_sim_MTM}
\end{figure}

In order to show how this nonequilibrium effect can strongly perturb the equilibrium stationary state of the system, \addeb{we plot on Fig.~\ref{fig:numerical_sim_MTM} the chemical potential obtained from direct numerical simulations of our model. The contact dynamics has been implemented numerically using $50$ links between the two systems,} with a typical exchange frequency at contact $\epsilon=0.01$ in order to guarantee an effective timescale separation between the bulk and the contact.
We used two systems of the same size ($|\Lambda|=10 000$ sites) in contact and keep the driving force of system $B$ equal to $f_{B}=2$. By symmetry, for $f_{A}=2$, the densities in each system should be the same, namely $\rho_{A}=\rho_{B}=\bar{\rho}=0.5$, as confirmed by numerical simulations (Fig.~\ref{fig:numerical_sim_MTM}). But when $f_{A}$ moves away from $f_{B}=2$, one can observe that the stationary density difference grows as well, leading to a significant effect. Also, as one can see on figure~\ref{fig:numerical_sim_MTM}, the agreement between theory and simulations is very good for this nonzero, but small, value of $\epsilon$. 

\subsection{Application to the KLS model}

As a further application of the framework expounded before, we now consider numerical simulations of a \comebb{well-known} lattice gas model, namely, the 2D KLS model \cite{katz1984nonequilibrium, zia2010twenty}. This system has already been considered in~\cite{pradhan2010nonequilibrium, pradhan2011approximate} to discuss a very similar situation of two systems brought into contact. We will discuss relations to our work in Sec.~\ref{sec:discussion} hereafter but we must as of now emphasize two major differences. First, we explicitly enforce a small exchange rate at contact so that Eq.~\eqref{eq:coarse_grained_transrate_rescaled_time} holds asymptotically. Second, we consider different dynamics at contact (namely, transition rates of Sasa-Tasaki and exponential type) enabling the factorization condition~\eqref{eq:factorization_micro_transrates} to hold.

Before \comebb{describing} our numerical simulation results, we briefly recall the dynamics of the KLS model and \comebb{introduce some} notations. We consider two lattices $\Lambda_{A}$ and $\Lambda_{B}$ in two dimensions with periodic boundary conditions. On each lattice, we call $n_{i}, \, i \in \Lambda_{k}$ ($k=A, \, B$) the occupation number and $\config_{k} = \{n_{i}\}_{i\in \Lambda_{k}}$ the whole configuration of system $k$. The energy of a configuration $\config$ reads as $H_{k}(\config_{k}) = - \tfrac{J_{k}}{2} \sum_{\langle i, j \rangle} n_{i}n_{j}$ where $\sum_{\langle i, j \rangle}$ refers to a sum on $(i, j)$ with $j$ nearest neighbor of $i$. We call $J_{k}$ the coupling constant setting the interactions between neighbors. Both systems are driven through the action of a homogeneous external force $f_{k}$ ($k=A,\, B$) along the $y$ direction.

As in~\cite{pradhan2010nonequilibrium, pradhan2011approximate}, we choose a continuous time asynchronous dynamics by moving one particle only at each time step. We assume local detailed balance and choose a Kawasaki rule for the transition rates in the bulk:
\begin{equation}
  \label{eq:glauber_tr_KLS}
  T(\config^{ij}|\config) = n_{i}(1-n_{j}) \frac{ \exp \Big(  \tfrac{\beta}{2} \left[ - \Delta H^{ij}(\config) + \bm{f}\cdot{}\bm{e}_{ij} \right] \Big) }{ \cosh  \Big(  \tfrac{\beta}{2} \left[ - \Delta H^{ij}(\config) + \bm{f}\cdot{}\bm{e}_{ij} \right] \Big) } \, .
\end{equation}
$\bm{e}_{ij}$ is the displacement vector such that $\bm{f}\cdot{}\bm{e}_{ij} = \pm f$ if the jump is along or opposite to the driving force $\bm{f}$, or $0$ if the latter is orthogonal to $\bm{f}$. $\config^{ij}$ refers to the configuration obtained from $\config$ by exchanging the occupation state of sites $i$ and $j$.
Since we consider periodic boundary conditions in \comebb{both directions}, systems $A$ and $B$ are brought into contact through a third dimension by allowing exchange of particles on few sites. In order to minimize potential effects of long-range correlations along the driving force, we place the contact sites on a same row along the $x$ axis. We consider two different dynamics at contact, namely, the exponential and the Sasa-Tasaki rules, which both obey local detailed balance and share the factorization property~\eqref{eq:factorization_micro_transrates}.

\subsubsection{Exponential rule}

In this subsection, we assume that the dynamics at contact is governed by the exponential rule that reads as 
\begin{align}
  \label{eq:exp_rule_KLS}
  & T_{c}(\config_{A}^{i_{A}-}, \config_{B}^{i_{B}+} | \config_{A}, \config_{B})  \\
  & \quad  = n_{i_{A}} ( 1 - n_{i_{B}} )  e^{-\tfrac{\beta}{2}\Delta H_{A}^{i_{A}-}(\config_{A})} e^{ -\tfrac{\beta}{2} \Delta H_{B}^{i_{B}+}(\config_{B})} \, ,\notag
\end{align}
for an exchange from $A$ to $B$ through the link $(i_{A}, i_{B})$. $\Delta H_{k}^{i_{k}\pm}(\config_{k})$ stands for the change of energy that follows the removal ($-$) or the addition ($+$) of one particle at site $i_{k}$ in system $k$. Exchanges from $B$ to $A$ can be easily recovered by swapping $n_{i_{A}}$ and $n_{i_{B}}$ as well as $+$ and $-$ signs.

From this exponential transition rate, one can derive the theoretical expression of chemical potentials according to the definition~\eqref{eq:def:chempot_contact} in the slow exchange limit. Indeed, the factors $\phi_{k}(\rho_{k}, \pm 1)$ of the macroscopic transition rates~\eqref{eq:factorization_transrate} read as
\begin{align}
  \label{eq:macro_tr_exp_KLS}
  \phi_{k}(\rho_{k}, +1) & = \sum_{i_{k} \in \Lambda_{k}^{c}} {\left\langle (1-n_{i_{k}}) e^{ -\tfrac{\beta}{2} \Delta H_{k}^{i_{k}+}(\config_{k})} \middle | \rho_{k} \right\rangle}_{k} \, , \\
  \phi_{k}(\rho_{k}, -1) & = \sum_{i_{k} \in \Lambda_{k}^{c}} {\left\langle n_{i_{k}} e^{ -\tfrac{\beta}{2} \Delta H_{k}^{i_{k}-}(\config_{k}) } \middle | \rho_{k} \right\rangle}_{k} \notag 
\end{align}

with $\Lambda_{k}^{c}$ the set of the sites involved in the contact and $\left\langle \cdot{} \middle | \rho_{k} \right\rangle_{k}$ the expectation with respect to the stationary probability distribution of isolated system $k$ at fixed density $\rho_{k}$.

Since we have considered periodic boundary conditions and a distribution of the contact sites orthogonal to the driving force, one can assume that all terms in the sums of Eqs.~\eqref{eq:macro_tr_exp_KLS} are equals. The chemical potential at contact hence reads
\begin{equation}
  \label{eq:chem_pot_exp_KLS}
  \mu_{k}^{\rm cont}(\rho_{k}) = \frac{ \left\langle n_{i_{k}} e^{ -\tfrac{\beta}{2} \Delta H_{k}^{i_{k}-}(\config_{k})} \middle | \rho_{k} \right\rangle_{k} }{ \left\langle (1-n_{i_{k}}) e^{ -\tfrac{\beta}{2} \Delta H_{k}^{i_{k}+}(\config_{k})} \middle | \rho_{k} \right\rangle_{k} } \, ,
\end{equation}
where $i_{k}$ can be any of the contact sites.

\begin{figure}[h]
\includegraphics[width=0.95\linewidth]{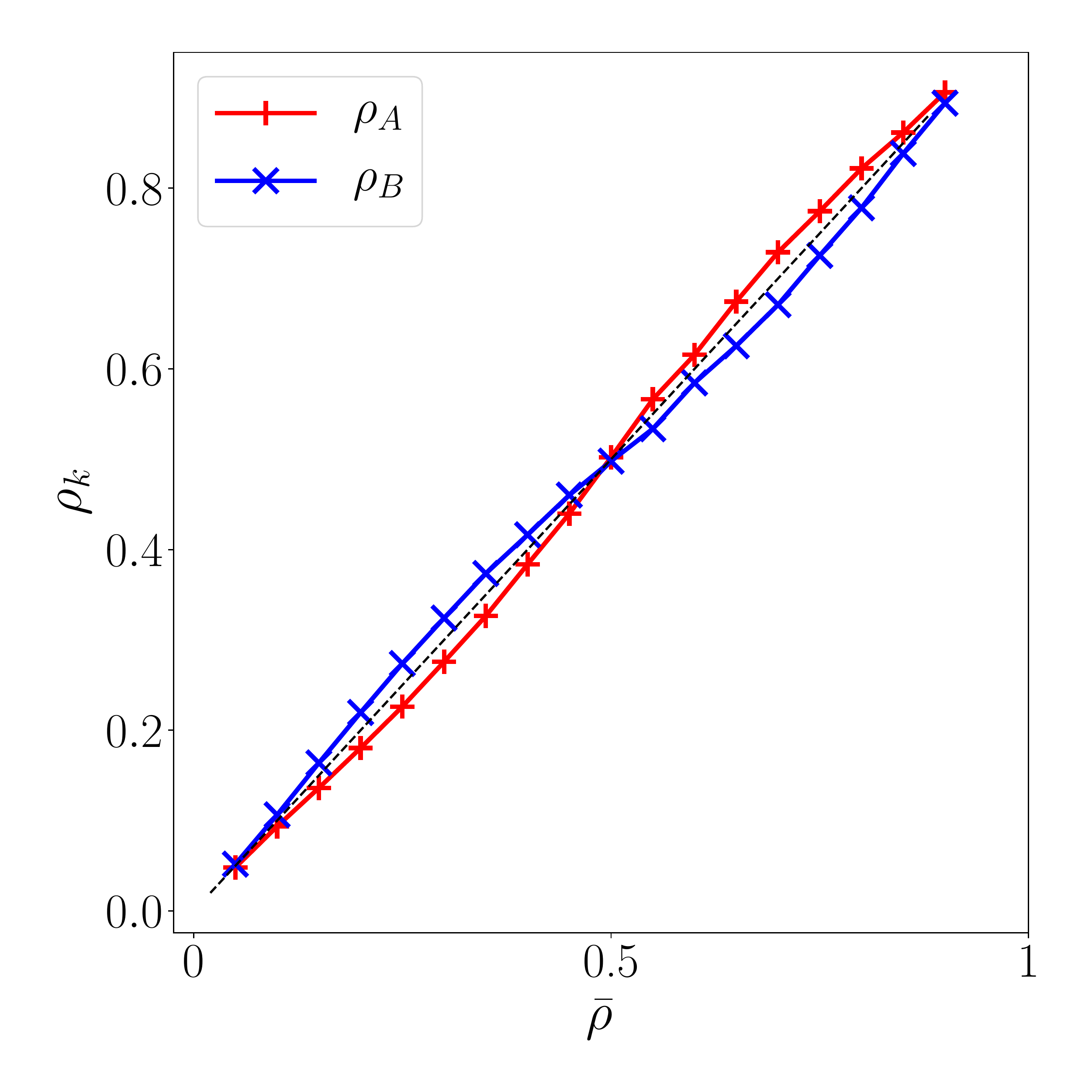}\\
\includegraphics[width=0.95\linewidth]{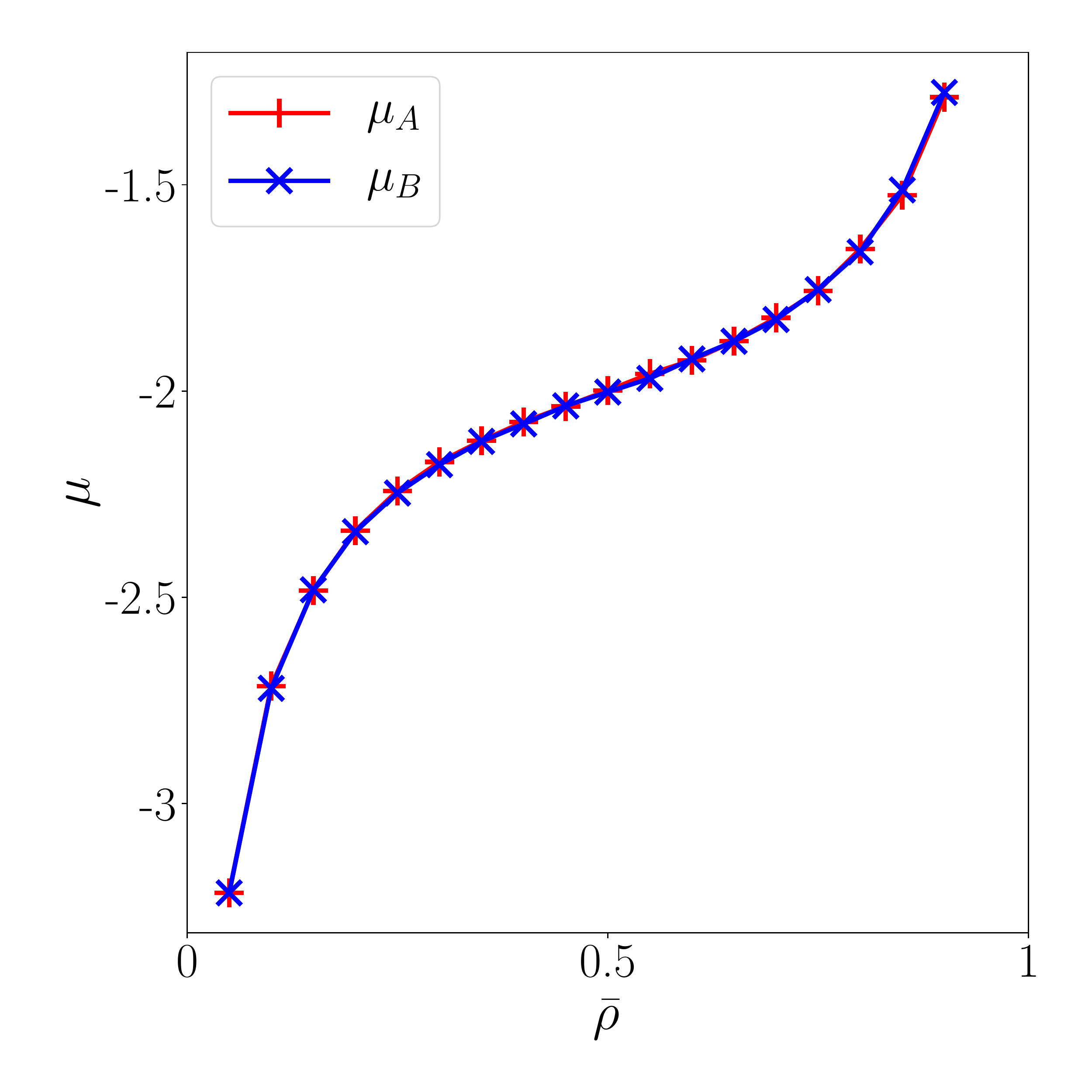}
  \caption{Numerical simulations of two KLS models A and B in contact at stationarity for different overall density, with an exponential dynamics at contact. Parameters are: $J_{A}=J_{B}=1$, $f_{A}=6$, $f_{B}=0$, $\beta=1$, $V_{A}=20\times 20$, $V_{B}=20\times 20$, $\epsilon=0.01$.
Top: Densities $\rho_A$ (red) and $\rho_B$ (blue) versus $\bar{\rho}$. Dashed line corresponds to the ideal equilibrium situation.
Bottom: Chemical potentials $\mu_{A}^{\rm cont}$ (red) and $\mu_{B}^{\rm cont}$ (blue) versus $\bar{\rho}$.}
\label{fig:numerical_KLS_exp}
\end{figure}

Figure~\ref{fig:numerical_KLS_exp} shows the balance of average densities and associated chemical potentials for two KLS systems $A$ and $B$ brought into contact for different overall densities $\bar{\rho}$. Both systems are of the same size $20\times 20$ and have the same coupling constant $J_{A}=J_{B}=1$. The system $A$ is forced with a driving force $f_{A}=6$ while system $B$ is kept at equilibrium ($f_{B}=0$). One can notice the quite important effect of the driving force $f_{A}$ (if both systems were in equilibrium, the densities of each of them would have been equal) \addjg{for intermediate densities (one does not expect any effect at low density for which interactions disappear as well as at high density for which the incompressibility wins (interactions being saturated))}. \addjg{In particular, one can observe that the effect of the external field is inverting around $\bar{\rho}\sim 0.5$.} But remarkably, the density shift is very well captured by the equalization of the chemical potentials $\mu_{A}^{\rm cont}$ and $\mu_{B}^{\rm cont}=\mu_{B}^{\rm eq}$ for this small but finite $\epsilon=0.01$.

\begin{figure}[h]
  \includegraphics[width=0.95\linewidth]{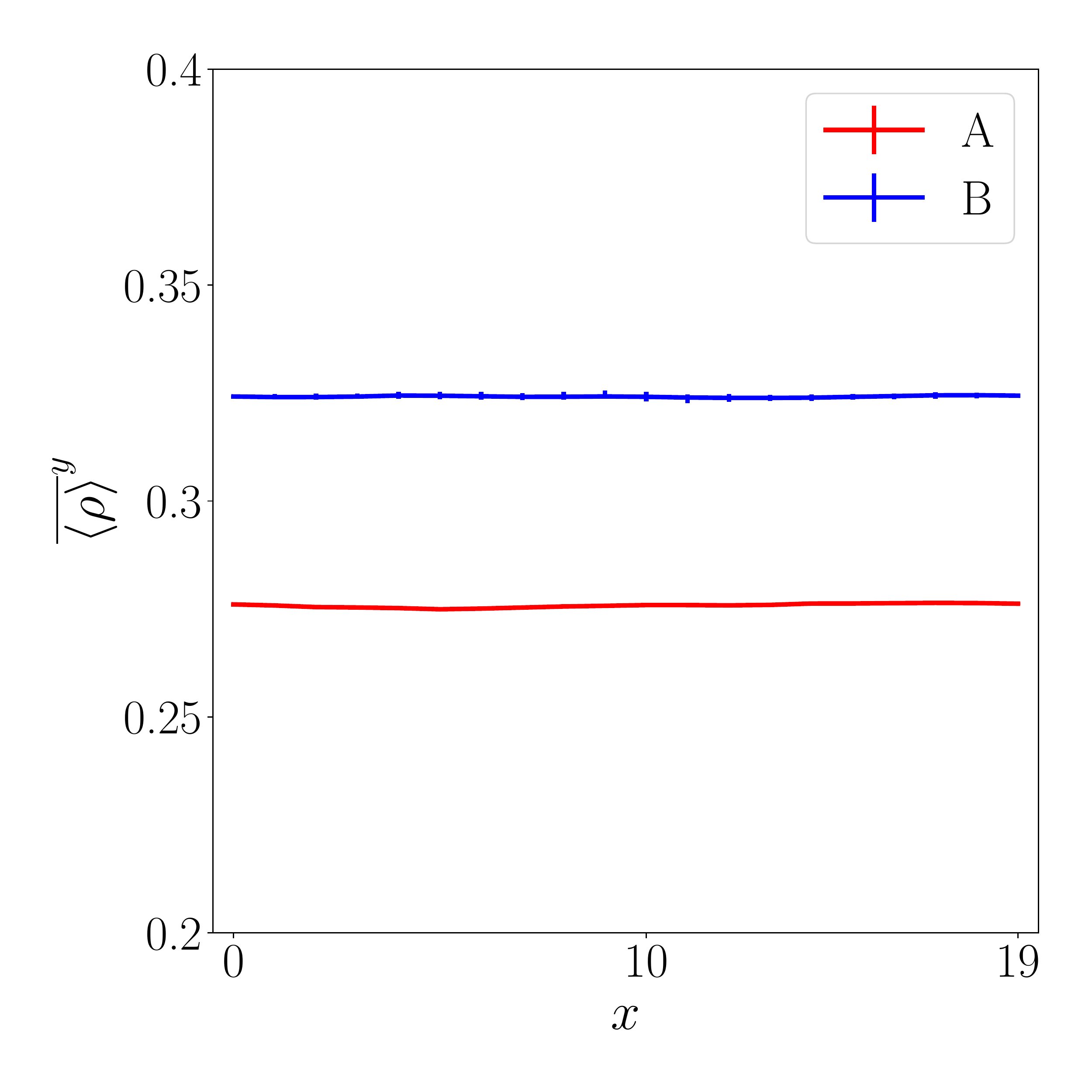}
  \caption{Average density profile along the $x$ direction for two KLS systems in contact. Plain line refers to the space average along $y$ and error bars indicate the maximum and minimum values along the $y$ direction. Dynamics is the exponential rule with following parameters: $\bar{\rho} = 0.3$, $J_{A}=J_{B}=1$, $f_{A}=6$, $f_{B}=0$, $\epsilon=0.01$. Red: system $A$. Blue: system $B$.}
\label{fig:density_profile_exp}
\end{figure}

\addjg{To support our assumption of homogeneity along the different sites at contact, we provide in Fig.~\ref{fig:density_profile_exp} the $y$-average stationary density profile along the $x$ direction; symbolically, $ \overline{ \langle \rho \rangle }^{y}(x) = \tfrac{1}{L_{y}}\int_{-L_{y}/2}^{L_{y}/2} \langle \rho \rangle (x, y) \mathrm{d}y$, for $\bar{\rho} = 0.3$, $J_{A}=J_{B}=1$, $f_{A}=6$, $f_{B}=0$, $\epsilon=0.01$. Error bars, barely visible, indicate the local minimum and maximum of average stationary density along the $y$ direction.}

In order to demonstrate the importance of the dynamics at contact, we now examine the same situation for which one has only replaced the exponential transition rates by Sasa-Tasaki ones. 

\subsubsection{Sasa-Tasaki rule}

For the Sasa-Tasaki rule, the transition rates read, using the same notation as above:
\begin{align}
  \label{eq:ST_rule_KLS}
  &  T_{c}(\config_{A}^{i_{A}-}, \config_{B}^{i_{B}+} | \config_{A}, \config_{B}) \\
  & \qquad = n_{i_{A}} ( 1 - n_{i_{B}} ) \exp\left(-\beta\Delta H_{A}^{i_{A}-}(\config_{A}) \right)  \notag \\
  & T_{c}(\config_{A}^{i_{A}+}, \config_{B}^{i_{B}-} | \config_{A}, \config_{B}) \\
  & \qquad = n_{i_{B}} ( 1 - n_{i_{A}} ) \exp\left(-\beta\Delta H_{B}^{i_{B}-}(\config_{B}) \right) \, . \notag
\end{align}

Computing the factors of the macroscopic transition rate in exactly the same way as in~\eqref{eq:macro_tr_exp_KLS} leads to
\begin{equation}
  \label{eq:chem_pot_st_KLS}
  \mu_{k}^{\rm cont}(\rho_{k}) = \frac{\left\langle n_{i_{k}} e^{-\beta \Delta H_{k}^{i_{k}-}(\config_{k})} \middle | \rho_{k} \right\rangle_{k}}{1 - \rho_{k}} \, .
\end{equation}

\begin{figure}[h]
\includegraphics[width=0.95\linewidth]{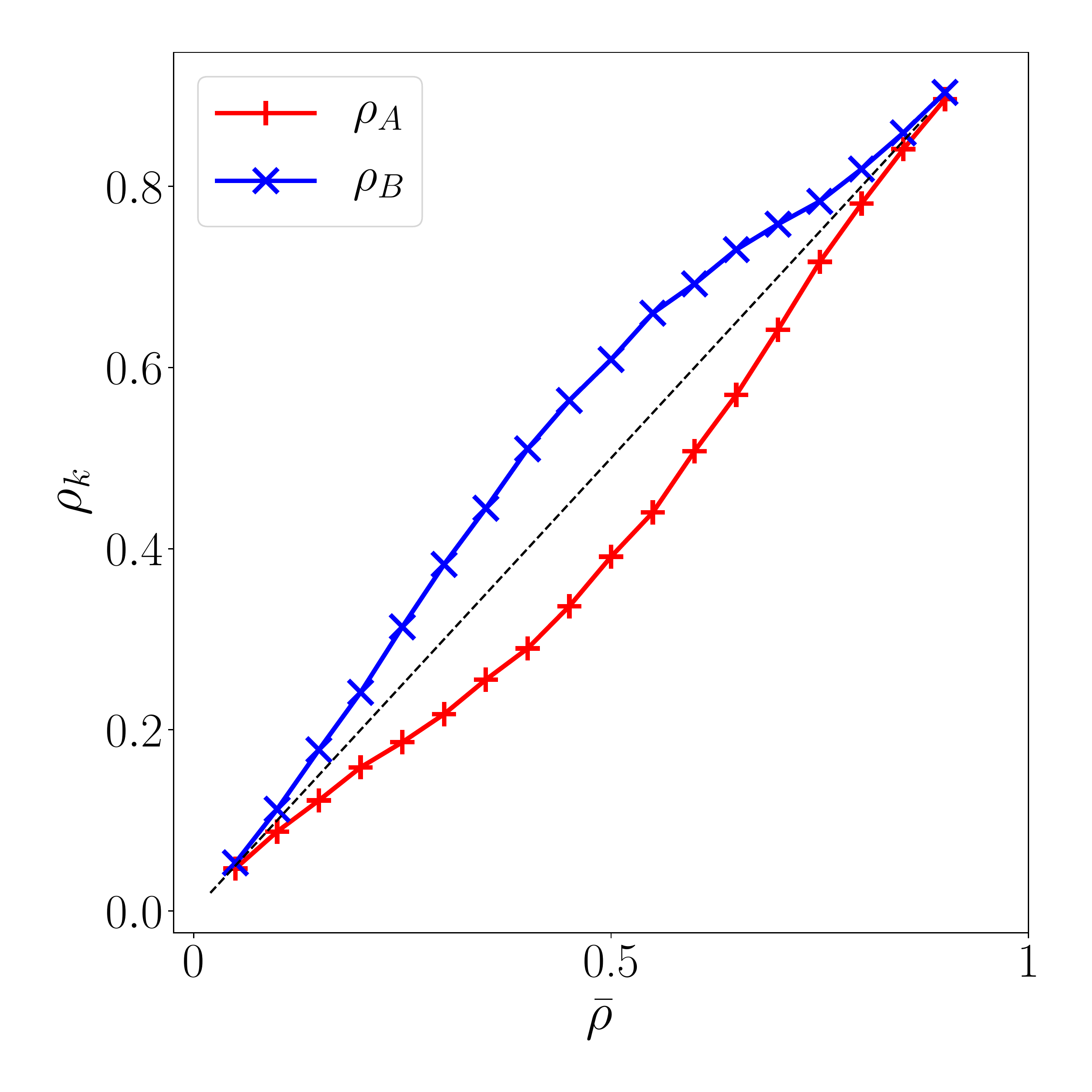}\\
\includegraphics[width=0.95\linewidth]{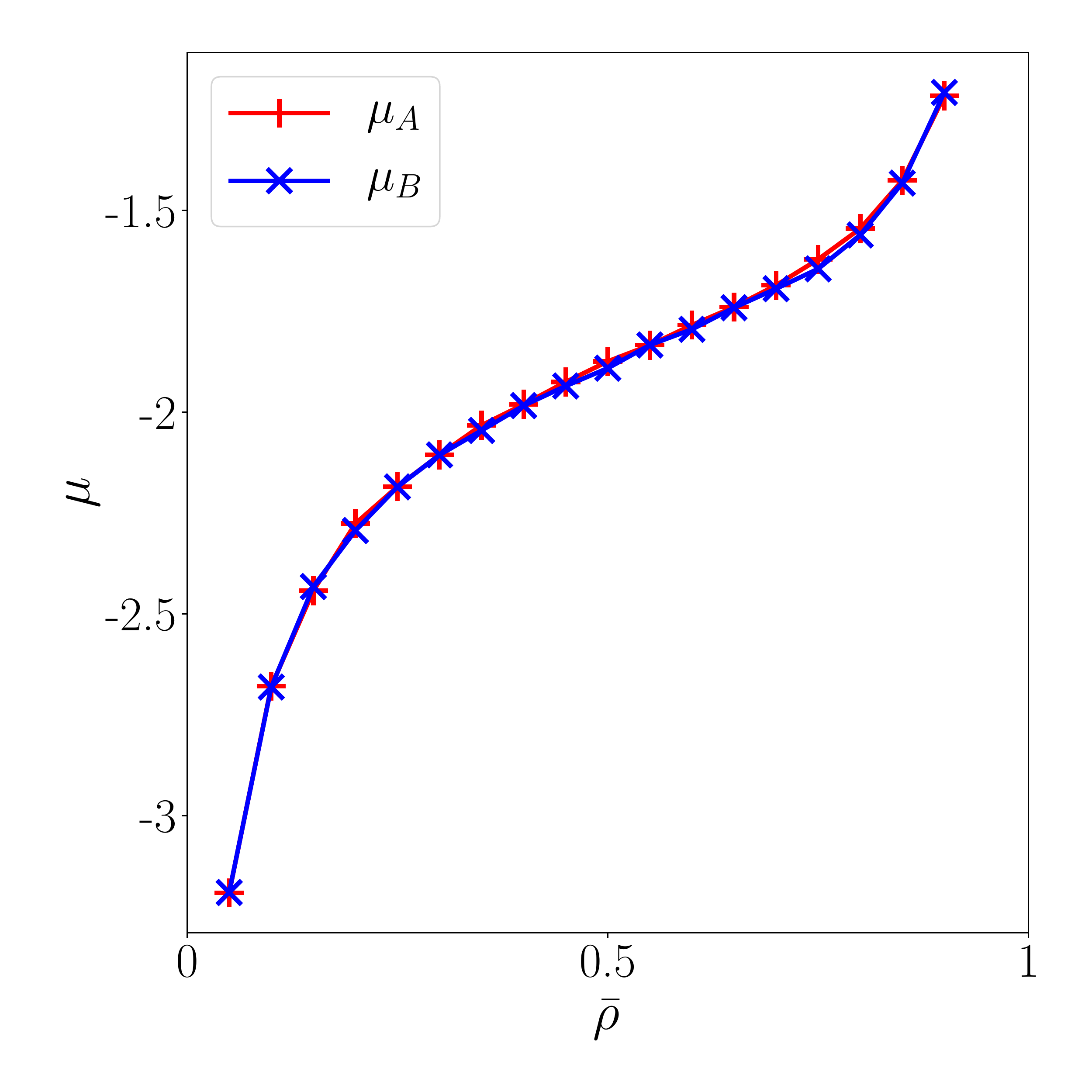}
  \caption{Numerical simulations of two KLS models $A$ and $B$ in contact at stationarity for different overall density, with a Sasa-Tasaki dynamics at contact. Parameters are: $J_{A}=J_{B}=1$, $f_{A}=6$, $f_{B}=0$, $\beta=1$, $V_{A}=20\times 20$, $V_{B}=20\times 20$, $\epsilon=0.01$.
Top: Densities $\rho_A$ (red) and $\rho_B$ (blue) versus $\bar{\rho}$. Dashed line corresponds to the ideal equilibrium situation.
Bottom: Chemical potentials $\mu_{A}^{\rm cont}$ (red) and $\mu_{B}^{\rm cont}$ (blue) versus $\bar{\rho}$.}
\label{fig:numerical_KLS_st}
\end{figure}

Figure~\ref{fig:numerical_KLS_st} is the analog of Fig.~\ref{fig:numerical_KLS_exp} for the Sasa-Tasaki rule. Comparison with the exponential rule shows that the Sasa-Tasaki dynamics has stronger impact on the density difference for the same driving force $f_{A}=6$. \addjg{Furthermore, one can observe that no inversion effect emerges here, the maximum impact being this time around $\rho \sim 0.5$.}

\begin{figure}[h]
  \includegraphics[width=0.95\linewidth]{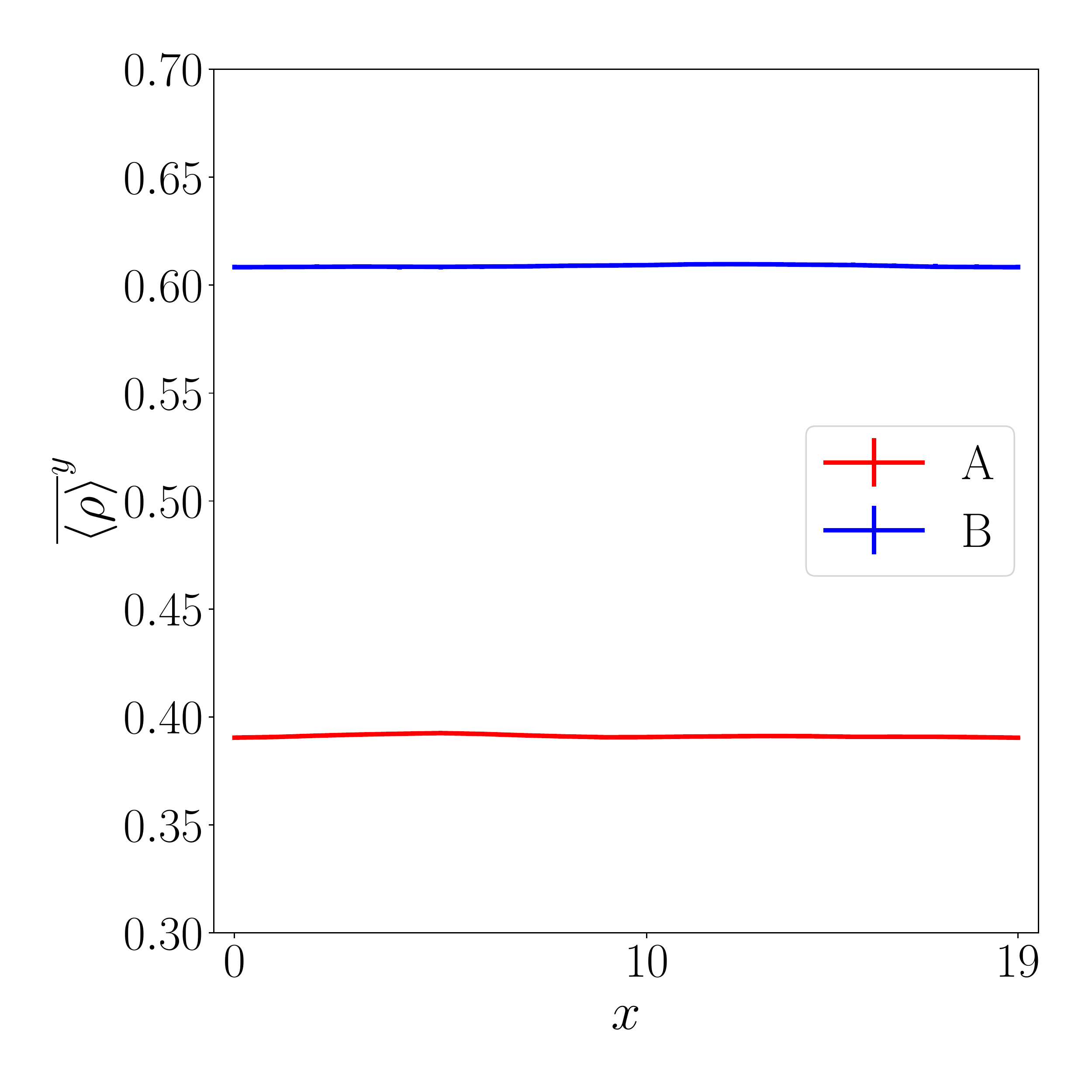}
  \caption{Average density profile along the $x$ direction. Plain line refers to the space average along $y$ and error bars indicate the maximum and minimum values along the $y$ direction. Dynamics is the Sasa-Tasaki rule with following parameters: $\bar{\rho} = 0.5$, $J_{A}=J_{B}=1$, $f_{A}=6$, $f_{B}=0$, $\epsilon=0.01$. Red: system $A$. Blue: system $B$.}
\label{fig:density_profile_st}
\end{figure}

\addjg{In a similar manner as the exponential rule, we provide in Fig.~\ref{fig:density_profile_st} estimation of the $y$-average density profile with respect to $x$ as well as the maxima and minima in the $y$ direction. One can observe that the density profile is rather flat everywhere, thus supporting the equivalence of all the sites belonging to the contact region.}

\section{Discussion}
~\label{sec:discussion}

In light of the general large-deviations framework and our previous examples of mass transport models, we now briefly discuss some of the main previous works
~\cite{pradhan2010nonequilibrium,pradhan2011approximate,chatterjee2015zeroth,dickman2014failure} closely related to the notion of out-of-equilibrium chemical potentials.

\subsection{Chemical potential and the zeroth law}

P. Pradhan \emph{et al.} \cite{pradhan2010nonequilibrium, pradhan2011approximate} discussed the existence of a thermodynamic structure with numerical simulations of two driven lattice gases in contact \cite{katz1984nonequilibrium}. 
The transition rates are of Metropolis type. Contrary to our setting, they have not assumed a vanishing exchange rate at contact. However, their measurement of the two-points correlation function across the contact (see Sec. V.B of~\cite{pradhan2011approximate}) shows that the latter is very small compared to the bulk correlations. This led them to assume the existence of a large-deviations principle for the probability distribution of densities with an additive large-deviations function, similar to the equilibrium case, but with chemical potentials in excess to account for the breaking of the zeroth law.

\addeb{Even though these numerical simulations were not done in the slow exchange limit, the observed breaking of the zeroth law for two driven lattice gases in contact is qualitatively} consistent with our general framework since the \comebb{zeroth law} is not expected to hold for most steady-state systems in contact. However, we would like to point out here that the assumption of the existence of a modified \emph{additive} large-deviations function is not consistent with the chosen transition rates at contact, namely the Metropolis rule. Indeed, the vanishing of the two-points correlation function across the contact suggests that the stationary probability distribution of the whole system may indeed be factorized as in \eqref{eq:coarse_grained_transrate_rescaled_time}. However, it can be shown that the Metropolis rule cannot lead to factorized coarse-grained transition rates (see \eqref{eq:coarse_grained_transrate_rescaled_time}) with the assumption of a factorized distribution. Since only one particle can be exchanged at a time, macroscopic detailed balance is expected to hold and thus leads to a nonadditive large-deviations function, according to \eqref{eq:macro_detailed_balance}.

This lack of additivity of the large-deviations function is also supported by the observed violation of the zeroth law when bringing a driven KLS system in contact with different equilibrium systems whose chemical potentials are known (see \cite{pradhan2011approximate}, Sec. III.B). Indeed, the chemical potential of the driven system measured through the ones of the equilibrium system is observed to depend on the coupling constant of the equilibrium systems, at odds with the equilibrium situation. This observation can be interpreted by noting that the nonadditive large-deviations function can be decomposed in a similar way as in Eq.~\eqref{eq:Iprime:Kawa:MTM}, namely
\begin{equation}
  \label{eq:non_additive_LD_excess}
  I'(\rho_{A}, \rho_{\rm eq}) = \mu^{\rm iso}(\rho_{A}) - \mu^{\rm eq}(\rho_{eq}) + \eta(\rho_{A}, \rho_{\rm eq}) \; .
\end{equation}
At stationarity, $\mu^{\rm iso}(\rho_{A}^{\ast}) + \eta(\rho_{A}^{\ast}, \rho_{\rm eq}^{\ast}) = \mu^{\rm eq}(\rho_{eq}^{\ast})$. Hence, measuring $\mu^{\rm eq}$ allows one to measure $\mu^{\rm iso} + \eta$ which depends on the parameters of both systems through the details of the nonfactorized microscopic transition rates in $\eta$ (see \eqref{eq:Iprime:Kawa:MTM} for such a dependence in another driven system). If $\eta$ were equal to zero, any change in the parameters of the equilibrium system would potentially modify the actual stationary densities $(\rho_{A}^{\ast}, \rho_{eq}^{\ast})$ but not the whole chemical potential function $\rho_{A}^{\ast} \to \mu^{\rm iso}(\rho_{A}^{\ast})$ since the latter should be independent of the equilibrium system. On the contrary, if $\eta \neq 0$ in \eqref{eq:non_additive_LD_excess}, any change in the parameters of the equilibrium system would modify both $\mu^{\rm eq}$ and $\eta$, then leading to different curves $\rho_{A}^{\ast} \to \mu^{\rm iso}(\rho_{A}^{\ast})+\eta(\rho_{A}^{\ast}, \rho_{\rm eq}^{\ast})$.

One should also point out that similar numerical results were found in \cite{dickman2014inconsistencies} for other driven lattice gases for which each site of both systems participates in the contact. The authors found that proper chemical potentials could be retrieved only in the small exchange rate limit with a factorized microscopic transition rates (the Sasa-Tasaki rule in this case), in agreement with the work presented here.

\subsection{Short-range correlations}

As an extension of the precursor analysis inspired by the ZRP \cite{bertin2006def, bertin2007intensive}, Chatterjee \emph{et al.} \cite{chatterjee2015zeroth} generalized the definition of chemical potentials for out-of-equilibrium steady-state systems in contact displaying short-range correlations in the bulks as well as in the contact area. Like our study, a slow exchange limit of mass at contact is assumed. This hypothesis allows one to write the stationary probability distribution as

\begin{align} \nonumber
P_{V}(\config_{A},\config_{B}|\bar{\rho}) &= \int d\rho_A d\rho_B \, P_{V}(\rho_{A}, \rho_{B}|\bar{\rho}) \\
& \qquad\qquad \times P_{V_{A}}(\config_{A}|\rho_{A}) P_{V_{B}}(\config_{B}|\rho_{B}) \,.
\label{eq:dist:PAB:discussion}
\end{align}

Hence, the absence of correlations between $A$ and $B$ lies in the factorization property of $P_{V}(\rho_{A},\rho_{B}|\bar{\rho})$, or, at large-deviations level, in the additivity property of $\mathcal{I}(\rho_{A}, \rho_{B}|\bar{\rho})$. They show, under the crucial assumption of short-range correlations inside each system (allowing factorization of the stationary distributions $P(\config_{k}|\rho_{k})$, $k=A,\,B$), that such an additivity property can hold only if microscopic detailed balance with respect to the nonequilibrium stationary distributions of both isolated systems, and then macroscopic detailed balance, is satisfied.

Although this study has mainly considered transition rates at contact that satisfy microscopic detailed balance with respect to the equilibrium distributions (and not the nonequilibrium distributions of isolated systems), 
\addeb{our analysis includes the situation} discussed in \cite{chatterjee2015zeroth} by simply assuming microscopic detailed balance with respect to the stationary nonequilibrium distributions of both isolated systems. 

We nevertheless point out that such an assumption appears less physically relevant if driven forces are orthogonal to the contact. 
\addeb{In particular, if the \addreview{microscopic} contact dynamics does not depend on the drive, microscopic detailed balance with respect to the distribution (\ref{eq:dist:PAB:discussion}) can only hold if the distributions $P_{V_{A}}(\config_{A}|\rho_{A})$ and $P_{V_{B}}(\config_{B}|\rho_{B})$ of the isolated systems do not depend on the drive, which is a very restrictive assumption. Alternatively, if these distributions depend on the drives, the contact dynamics has to be fine tuned with the drives for the microscopic detailed balance conditions to hold.}

As for the short-range correlation assumption, we agree that \addeb{the hypothesis made in \cite{chatterjee2015zeroth} allows one to justify the existence of chemical potentials for isolated systems, consistently with the discussion in Sec.~\ref{sec:sub:relation_mucont_muiso}.
However, our study provides general conditions to reach the additivity property as discussed in Sec.~\ref{sec:additivity_large_dev}, a particular case of which yields back the condition assumed in \cite{chatterjee2015zeroth}.}

We note that with a slight modification of microscopic dynamics at contact, our study generalizes the work of \cite{chatterjee2015zeroth} by considering more general \emph{factorized} dynamics at contact, leading to a broader form of additivity for which chemical potentials at contact and of isolated systems do not necessarily coincide. 


\subsection{Position of the contact in multidimensional systems}

Eventually, we discuss briefly the results expounded in \cite{dickman2014failure} (see also \cite{dickman2014inconsistencies}) where the author discusses the effect of the position and the extension of the contact region. For the same factorized microscopic transition rates, the author indeed showed that different final densities could be reached simply by modifying the position of the contact (pointwise in the bulk or near the boundaries, along the edges of both systems, etc.). This effect can be easily interpreted in our framework. Indeed, the macroscopic transition rates are averages of the microscopic ones with respect to the stationary distribution of the configurations in the contact area. The way in which the latter differs from the equilibrium distribution generically depends on the position of the contact area with respect to the bulk of the systems in contact. In particular, as shown in \cite{dickman2014failure}, perturbations near boundaries do modify the chemical potentials at contact (when additivity holds), thus leading to different stationary densities in both systems.

\section{Conclusion}

In this paper, we have shown how a nonequilibrium chemical potential can be defined for two driven systems in slow exchange limit at contact. This definition relies on the additivity property of the large-deviations function describing the statistics of the densities of the systems in contact.
A sufficient condition for the additivity property to hold is that the coarse-grained dynamics of the exchange of mass satisfies a detailed balance property, and that the microscopic dynamics of the contact factorizes with respect to the two systems.

Quite importantly, the nonequilibrium chemical potential of the two systems in contact lacks an equation of state, and explicitly depends on the contact dynamics (see also \cite{guioth2019lack} for a similar result in the framework of gases of active particles).
As a consequence, the steady-state densities of the two systems also depend on the contact dynamics, even in the slow exchange limit at contact considered here.
However, the zeroth law of thermodynamics still holds, but only for restricted classes of systems defined by including (half of) the contact dynamics in the definition of the system. We have also shown that the chemical potential of systems in contact differs from that of the isolated systems, and can be reexpressed by introducing a deviation with respect to a reference state: either the equilibrium state or the isolated driven system.

We have also discussed our results on the explicit example of an exactly solvable driven lattice gas, and shown on the example of the KLS model that the method also applies to models with unknown steady-state distributions, using a numerical procedure to determine the chemical potentials.

Future work may follow, among others, two research lines.
First, it could be of interest to explore the corrections to the slow exchange limit at contact for a small but finite exchange rate. Calculations are much harder in this case, but preliminary results suggest that the additivity of the large-deviations function is generically broken in such a situation,
\addeb{which may lead to further interesting effects. For instance,
unequal steady-state densities have been found recently in zero-range processes in contact \cite{Dickman19uphill}, while the slow exchange limit predicts equal densities in such models where the probability distributions of isolated systems do not depend on the drives.}
Hence it is likely that no chemical potential can be defined beyond the slow exchange limit at contact.
A second, and perhaps more promising, line of research would be to extend the large-deviations approach to evaluate the large-deviations functional of the full density profile, in the spirit of macroscopic fluctuation theory \cite{bertini2015macroscopic}.
One of the goals of such an extension would be to deal with smooth interfaces at the contact between the two systems, instead of sharp ones as considered in the present work.
An extension along this line may be of interest to describe for instance phase coexistence in active systems, perhaps providing some support to recent phenomenological approaches aimed at describing this phenomenon \cite{solon2018generalized-pre,solon2018generalized-njp}.
One may draw inspiration from the method presented in \cite{Barre2015} to evaluate the large-deviations functional of the density profile in systems of active Brownian particles.


\begin{acknowledgments}
J.G. acknowledges fundings from the Royal Society and the French Ministry of Higher Education and Research. 
\end{acknowledgments}


\appendix

\section{Stationary state: link between the vanishing of $I(\rho_{A} | \bar{\rho})$ and the current $J(\rho_{A})$}
\label{sec:appA}

To show explicitly that the stationary state $\rho_{A}^{\ast}$ is the minimum of $I(\rho_{A}|\bar{\rho})$ and makes the current $J$ vanish, we use, following \cite{ge2017mathematical}, the stationary Hamilton-Jacobi equation \eqref{eq:stationary_HJ_eq_rhoA} evaluated along the deterministic path obeying \eqref{eq:def:determinist_current}. For $\rho_{A}(t)$ solution of \eqref{eq:def:determinist_current}, one has
  \begin{align}
    \label{eq:I_SS-liapunov_function}
    \deriv{I(\rho_{A}(t)|\bar{\rho})}{t} & = J(\rho_{A}(t))I^{\prime}(\rho_{A}(t)|\bar{\rho}) \\
                                         & = \sum_{\Delta N_{A} \neq 0} \varphi(\rho_{A}(t),\Delta N_{A}) \Delta N_{A} I^{\prime}(\rho_{A}(t)|\bar{\rho}) \, .  \notag 
   \end{align}
Let us emphasize that $I(\rho_{A})$ is the stationary large-deviations function, solution of \eqref{eq:stationary_HJ_eq_rhoA}. Hence the only time dependent quantity is the average density $\rho_A(t)$.

Since the inequality $e^{x}-1 \geqslant x$ holds for all $x$ with equality only when $x=0$, $\Delta N_{A}I'(\rho_{A}(t)|\bar{\rho}) \leqslant e^{\Delta N_{A}I'(\rho_{A}(t)|\bar{\rho})}-1$, the last equality in \eqref{eq:I_SS-liapunov_function} yields
  \begin{align}
    \label{eq:I_liapunov_function_bis}
    &\deriv{I(\rho_{A}(t)|\bar{\rho})}{t} \\
    &\hspace{1.5em} \leqslant \sum_{\Delta N_{A}} \varphi(\rho_{A}(t),\Delta N_{A}) \left(e^{\Delta N_{A} I'(\rho_{A}(t)|\bar{\rho})} - 1 \right) = 0 ,  \notag 
  \end{align}
  because the last term is the left-hand side of the Hamilton-Jacobi equation \eqref{eq:stationary_HJ_eq_rhoA}. The stationary large-deviations function $I(\rho_{A}|\bar{\rho})$ thus plays the role of a \emph{Lyapunov function} for the macroscopic dynamics. This implies that a steady-state $\mathrm{d}I(\rho_{A}(t)|\bar{\rho})/\mathrm{d}t = 0$ corresponds to $\rho_A(t)=\rho_A^{\ast}$ with $I'(\rho_{A}^{\ast}|\bar{\rho}) = 0$.
So if $J(\rho_{A}^{\ast}) = 0$, Eqs.~(\ref{eq:I_SS-liapunov_function}) and (\ref{eq:I_liapunov_function_bis}) imply that $I'(\rho_{A}^{\ast}|\bar{\rho}) = 0$.

Let us now show that, conversely, $I'(\rho_{A}^{\ast}|\bar{\rho}) = 0$ implies $J(\rho_{A}^{\ast}) = 0$.
Taking the derivative with respect to $\rho_{A}$ of the stationary Hamilton-Jacobi equation \eqref{eq:stationary_HJ_eq_rhoA} leads to
  \begin{multline}
    \label{eq:derivative_HJ_rhoA}
    0 = \sum_{\Delta N_{A}} \deriv{\varphi}{\rho_{A}}(\rho_{A}, \Delta N_{A})\left(e^{\Delta N_{A} I'(\rho_{A}|\bar{\rho})} - 1 \right) \\
       + I''(\rho_{A}|\bar{\rho}) \, \sum_{\Delta N_{A}} \varphi(\rho_{A},\Delta N_{A}) \, \Delta N_{A}  \, e^{\Delta N_{A} I'(\rho_{A}|\bar{\rho})} \; .
  \end{multline}
  At the stationary point $\rho_{A}^{\ast}$ for which $I'(\rho_{A}^{\ast}|\bar{\rho})=0$, the last equation reads
  \begin{equation}
    \label{eq:second_derivative_I_SS}
    J(\rho_{A}^{\ast}) I^{\prime\prime}(\rho_{A}^{\ast}) = 0 \; .
  \end{equation}
  Then, if $I^{\prime\prime}(\rho_{A}^{\ast}|\bar{\rho}) \neq 0$ (actually, $I^{\prime\prime}(\rho_{A}^{\ast}|\bar{\rho}) > 0$ to ensure convexity), $I^{\prime}(\rho_{A}^{\ast}|\bar{\rho}) = 0$ implies $J(\rho_{A}^{\ast}) = 0$.
We have thus shown the equivalence between the properties
$J(\rho_{A}^{\ast}) = 0$ and $I'(\rho_{A}^{\ast}|\bar{\rho}) = 0$.

\section{Exactly solvable lattice model}
\label{sec:appB}

\subsubsection*{Isolated chemical potential}

Even if the distribution $P$ is not factorized, one can define a chemical potential associated with one isolated system, related to the partition function $Z(|\Lambda|, N)$. Indeed, from the normalization of the stationary probability \eqref{eq:stationary_probability_MTM}, the partition function of our model reads as
\begin{equation}
  \label{eq:partition_function_MTM}
  Z(|\Lambda|, N) = 2 \sum_{\{n_{i}\}_{i\in\Lambda}}\left[ \, \prod_{k=1}^{L} f_{+}(n_{2k})f_{-}(n_{2k+1}) \, \right] \delta_{\, \sum_{i\in\Lambda}n_{i}, \, N}
\end{equation}
where $f_{+}(n) = \exp\left[-\varepsilon(n) + fn\right]$ and $f_{-}(n) = \exp\left[-\varepsilon(n) - fn\right]$. Introducing the Fourier transform of the Kronecker delta, one obtains
\begin{equation}
  \label{eq:partition_function_MTM_Fourier}
  Z(|\Lambda|, N) = 2 \int_{-\pi}^{\pi}\! \ddr\theta e^{-|\Lambda|\left[i\theta \bar{\rho} - \tfrac{1}{2} \ln(z_{+}(i\theta) z_{-}(i\theta))\right]} \, ,
\end{equation}
where one has introduced $z_{\alpha}(x) = \sum_{n}f_{\alpha}(n)e^{xn}$, $\alpha=\pm$ and $\bar{\rho}=N/|\Lambda|$. Assuming that there is only one saddle-point at $\mu(\bar{\rho})$, the partition function eventually reads
\begin{equation}
  \label{eq:partition_function_MTM_LDev}
  Z(|\Lambda|, N) \asymp  e^{-|\Lambda|\left[ \mu(\bar{\rho})\bar{\rho} - \tfrac{1}{2}\ln(z_{+}(\mu(\bar{\rho}))z_{-}(\mu(\bar{\rho}))) \right]} 
\end{equation}
with the implicit equation verified by $\mu(\bar{\rho})$ reading
\begin{equation}
  \label{eq:mu_iso_implicit_def}
  \bar{\rho} = \frac{1}{2}\left(\frac{z_{+}^{\prime}(\mu(\bar{\rho}))}{z_{+}(\mu(\bar{\rho}))} + \frac{z_{-}^{\prime}(\mu(\bar{\rho}))}{z_{-}(\mu(\bar{\rho}))} \right) \; .
\end{equation}
The quantity $\mu(\bar{\rho})$ is naturally interpreted as the chemical potential associated with the isolated system.

\subsubsection*{Single site marginal distribution}

Integrating over all except one site the stationary distribution \eqref{eq:stationary_probability_MTM}, the single site probability distribution reads
\begin{align}
  \label{eq:single_site_probability_distribution_MTM}
  P(n|\bar{\rho}) & = \frac{\exp\left[\mu(\bar{\rho})n\right]}{2}\left( \frac{f_{+}(n)}{z_{+}(\mu(\bar{\rho}))} + \frac{f_{-}(n)}{z_{-}(\mu(\bar{\rho}))} \right) \\ 
                  & =  \frac{\exp\left[\mu(\bar{\rho})n -\varepsilon(n)\right]}{z_{0}(\mu(\bar{\rho}))} \exp\left(\upsilon[\mu, f](n)\right), \notag
\end{align}
with
\begin{equation}
  \label{def:eq:upsilon:mu:f}
  \exp\left(\upsilon[\mu,f](n)\right)  = \frac{z_{0}(\mu)}{2}  \left( \frac{e^{fn}}{z_{+}(\mu)} + \frac{e^{-fn}}{z_{-}(\mu)} \right) .
\end{equation}
The quantity $z_{0}(x)$ reads $\sum_{n}f_{0}(n)e^{xn}$, where $f_{0}(n)=\exp\left[-\varepsilon(n)\right]$, \emph{i.e.}, the stationary weight for the driving force $f=0$.

\subsubsection*{Detailed computation of the chemical potentials for the natural and the Sasa-Tasaki dynamics}

\paragraph*{Natural dynamics.} For the natural dynamics, the explicit expressions of the factors $\phi_A(\rho_{A},\Delta N_A)$ and $\phi_B(\rho_{B},\Delta N_B)$ are given by
\begin{align}
  \label{eq:transrate_natural_dyn_1part_MTM}
  \phi_A(\rho_{A}, +1) & = \sum_{n_{i_{A}}=0}^{n_{\mathrm{max}}^{A}-1} e^{-\varepsilon_{A}(n_{i_{A}}+1)} P(n_{i_{A}}|\rho_{A}) \\
  \phi_A(\rho_{A}, -1) & = \sum_{n_{i_{A}}=1}^{n_{\mathrm{max}}^{A}} e^{-\varepsilon_{A}(n_{i_{A}}-1)}P(n_{i_{A}}|\rho_{A}) \notag\\
  \phi_B(\rho_{B},+1) & = \sum_{n_{j_{B}}=0}^{n_{\mathrm{max}}^{B}-1} e^{-\varepsilon_{B}(n_{j_{B}}+1)}P(n_{j_{B}}|\rho_{B}) \notag\\
  \phi_B(\rho_{B},-1) & = \sum_{n_{j_{B}}=1}^{n_{\mathrm{max}}^{B}} e^{-\varepsilon_{B}(n_{j_{B}}-1)}P(n_{j_{B}}|\rho_{B}) \notag
\end{align}
with $|\Lambda_{B}|\rho_{B}= N - |\Lambda_{A}|\rho_{A}$. 
Microscopic transition rates being factorized, chemical potentials associated with each system can be defined and read, according to \eqref{eq:def:chempot_contact},
\begin{equation}
  \label{eq:chem_pot_natural_dyn_MTM}
  \mu_{k}^{\mathrm{cont}}(\rho_{k}) = \ln \frac{\sum_{n_{k}=0}^{n_{\mathrm{max}}^{k}-1} e^{-\varepsilon_{k}(n_{k})}P(n_{k}+1|\rho_{k})}{\sum_{n_{k}=0}^{n_{\mathrm{max}}^{k}-1} e^{-\varepsilon_{k}(n_{k}+1)}P(n_{k}|\rho_{k})} \; ,
\end{equation}
where $k=A,\, B$. Using the expression of the single site probability distribution given in Eq.~\eqref{eq:single_site_probability_distribution_MTM}, one finally obtains Eq.~\eqref{eq:excess_chem_pot_ND_ST_MTM}.

\paragraph*{Sasa-Tasaki rule.} For the Sasa-Tasaki dynamics rule \eqref{eq:ST_rule_contact_MTM}, the coarse-grained transition rates are also factorized, but the expressions of the factors $\phi_A(\rho_{A},\Delta N_A)$ and $\phi_B(\rho_{B},\Delta N_B)$ differ from that of the natural dynamics.
They read as
\begin{align}
  \label{eq:transrate_ST_dyn_1part_MTM}
  \phi_A(\rho_{A}, +1) & = \sum_{n_{i_{A}}=0}^{n_{\mathrm{max}}^{A}-1} P(n_{i_{A}}|\rho_{A}) \, , \\
  \phi_A(\rho_{A}, -1) & = \sum_{n_{i_{A}}=1}^{n_{\mathrm{max}}^{A}} e^{-\left[\varepsilon_{A}(n_{i_{A}}-1)-\varepsilon_{A}(n_{i_{A}})\right]}P(n_{i_{A}}|\rho_{A}) \notag \, ,\\
  \phi_B(\rho_{B}, +1) & = \sum_{n_{j_{B}}=0}^{n_{\mathrm{max}}^{B}-1}P(n_{j_{B}}|\rho_{B}) \notag \, ,\\
  \phi_B(\rho_{B}, -1) & = \sum_{n_{j_{B}}=1}^{n_{\mathrm{max}}^{B}} e^{-\left[\varepsilon_{B}(n_{j_{B}}-1) - \varepsilon_{B}(n_{j_{B}}) \right] }P(n_{j_{B}}|\rho_{B}) \notag
\end{align}
with again $|\Lambda_{B}|\rho_{B}= N - |\Lambda_{A}|\rho_{A}$. The exclusion rule, i.e., the fact that there can be at most $n_{\mathrm{max}}$ particles on a single site, generates a dependence on the recipient system although the transition rates only involve the energy variation of the sender system. The term reminiscent of the exclusion rule reads as
\begin{equation}
\sum_{n_{k}=0}^{n_{\mathrm{max}}^{k} - 1} P(n_{k}|\rho_{k}) = 1 - P(n_{\max}^{k}|\rho_{k})
\end{equation}
by normalization. When $n_{\mathrm{max}}\to\infty$, one expects $P(n_{\mathrm{max}}^{k}|\rho_{k})\to 0$ and this extra dependence vanishes. Apart from this remark, calculations remain qualitatively similar to the previous case and chemical potentials read as
\begin{equation}
  \label{eq:chem_pot_ST_MTM}
  \mu_{k}^{\mathrm{cont}}(\rho_{k}) = \ln \frac{\sum_{n_{i_{k}}=1}^{n_{\mathrm{max}}^{k}} e^{-\left[\varepsilon_{k}(n_{i_{k}}-1)-\varepsilon_{k}(n_{i_{k}})\right]}P(n_{i_{k}}|\rho_{k})}{1-P(n_{\mathrm{max}}^{k}|\rho_{k})} \; .
\end{equation}
\comebb{Similarly to the case of} the natural dynamics, one can use the single-site probability distribution \eqref{eq:single_site_probability_distribution_MTM} given above to write $\mu_{k}^{\rm cont}$ in terms of $\mu_{k}^{\mathrm{iso}}$ and $\eta_{k}^{\rm (ST)}$ displayed on \eqref{eq:excess_chem_pot_ND_ST_MTM}.

\bibliography{biblio_contact_lattgas}

\end{document}